\documentclass[
reprint,
superscriptaddress,
nofootinbib,
amsmath,amssymb,
aps,
prx,
]{revtex4-2}
\usepackage{array}
\usepackage{amsmath}
\usepackage{bm}
\usepackage{mathtools}
\usepackage{mathrsfs}
\usepackage{physics}
\usepackage{graphicx}
\usepackage{here}
\usepackage{dcolumn}
\usepackage{makecell}
\usepackage{hyperref}
\usepackage{xcolor}
\usepackage{ascmac}
\usepackage{natbib}
\usepackage{appendix}

\hypersetup{
setpagesize=false,
bookmarksnumbered=true,
bookmarksopen=true,
colorlinks=true,
linkcolor=blue,
citecolor=blue,
urlcolor=blue
}

\newcommand{\bma}{\bm{a}}
\newcommand{\bmg}{\bm{g}}
\newcommand{\bmp}{\bm{p}}
\newcommand{\bmr}{\bm{r}}

\newcommand{\bmv}{\bm{v}}
\newcommand{\bmA}{\bm{A}}
\newcommand{\bmF}{\bm{F}}
\newcommand{\bmlambda}{\bm{\lambda}}
\newcommand{\bmmu}{\bm{\mu}}
\newcommand{\bmnu}{\bm{\nu}}

\newcommand{\mcF}{\mathcal{F}}
\newcommand{\mcL}{\mathcal{L}}
\newcommand{\mcT}{\mathcal{T}}
\newcommand{\mcW}{\mathcal{W}}
\newcommand{\bmmcT}{\bm{\mathcal{T}}}
\newcommand{\opt}{\mathrm{opt}}
\newcommand{\ini}{\mathrm{ini}}
\newcommand{\fin}{\mathrm{fin}}
\colorlet{MYBLUE}{blue}

\begin{document}

\preprint{APS/123-QED}

\title{Optimizing optimal transport: Role of final distributions in \texorpdfstring{\\ finite-time thermodynamics}{finite-time thermodynamics}}

\makeatletter
\author{Kaito Tojo\frontmatter@footnote{These authors contributed equally to this work}}
\email{tojo@noneq.t.u-tokyo.ac.jp}
\affiliation{
  Department of Applied Physics, The University of Tokyo, 7-3-1 Hongo, Bunkyo-ku, Tokyo 113-8656, Japan
}

\author{Rihito Nagase\frontmatter@footnote{These authors contributed equally to this work}}
\affiliation{
  Department of Applied Physics, The University of Tokyo, 7-3-1 Hongo, Bunkyo-ku, Tokyo 113-8656, Japan
}

\author{Ken Funo}
\affiliation{
  Department of Applied Physics, The University of Tokyo, 7-3-1 Hongo, Bunkyo-ku, Tokyo 113-8656, Japan
}
\author{Takahiro Sagawa}
\affiliation{
  Department of Applied Physics, The University of Tokyo, 7-3-1 Hongo, Bunkyo-ku, Tokyo 113-8656, Japan
}
\affiliation{
    Quantum-Phase Electronics Center (QPEC), The University of Tokyo, 7-3-1 Hongo, Bunkyo-ku, Tokyo 113-8656, Japan
}
\affiliation{
    Inamori Research Institute for Science (InaRIS), Kyoto-shi, Kyoto 600-8411, Japan
}

\begin{abstract}
    Performing thermodynamic tasks within finite time while minimizing thermodynamic costs is a central challenge in stochastic thermodynamics.
    Here, we develop a unified framework for optimizing the thermodynamic cost of performing various tasks in finite time for overdamped Langevin systems.
    Conventional optimization of thermodynamic cost based on optimal transport theory leaves room for varying the final distributions according to the intended task, enabling further optimization.
    Taking advantage of this freedom, we use Lagrange multipliers to derive the optimal final distribution that minimizes the thermodynamic cost.
    Our framework applies to a wide range of thermodynamic tasks, including particle transport, thermal squeezing,
    and information processing such as information erasure, measurement, and feedback.
    Our results are expected to provide design principles for information-processing devices and thermodynamic machines that operate at high speed with low energetic costs.
\end{abstract}

\maketitle

\section{Introduction}\label{sec:introduction}
Optimal transport theory~\cite{Villani_optimal_transport,Santambrogio_Optimal_Transport, MAL-073} is a powerful mathematical methodology that has developed from the optimization problem of how to transfer one probability distribution into another with the smallest possible transport cost.
Solving such a transport problem yields an optimal transport plan and a distance between probability distributions known as the Wasserstein distance.
In recent years, optimal transport theory has been applied across diverse fields, including deep learning~\cite{pmlr-v70-arjovsky17a}, image processing~\cite{Rubner:2000aa}, and molecular biology~\cite{SCHIEBINGER2019928}.

In the context of thermodynamics, the connections between optimal transport theory and stochastic thermodynamics~\cite{Sekimoto_Stochastic_Energetics, Seifert_2012, Peliti_Pigolotti_Stochastic_Thermodynamics,Shiraishi_Stochastic_Thermodynamics, PhysRevX.7.021051} have been investigated.
The lower bound on the entropy production set by the second law of thermodynamics is achievable only in the quasistatic limit, which makes it unattainable in finite time~\cite{PhysRevLett.117.190601}.
Therefore, tighter bounds than the second law have been pursued theoretically~\cite{PhysRevLett.95.190602, Schmiedl_2008, PhysRevLett.105.150603, PhysRevLett.98.108301, PhysRevLett.99.100602, PhysRevLett.108.190602, Blaber_2023, PhysRevLett.121.070601, PhysRevLett.121.030605, PhysRevX.10.021056, PhysRevLett.117.190601, PhysRevLett.114.158101, PhysRevLett.116.120601, Horowitz:2020aa, doi:10.1073/pnas.2321112121}
and experimentally~\cite{PhysRevLett.126.170601, PhysRevLett.129.270601, doi:10.1073/pnas.2301742120, PhysRevE.111.044114, PhysRevLett.125.210601, PhysRevX.14.021032, Oikawa:2025aa, PhysRevResearch.2.022044}.
Specifically, optimal transport theory yields the fundamental achievable bound and the explicit optimal protocol in finite time.
Such optimal speed limits have been formulated for overdamped Langevin systems~\cite{PhysRevLett.106.250601, Aurell:2012aa, dechant2019thermodynamicinterpretationwassersteindistance, PhysRevResearch.3.043093, PhysRevLett.125.100602, PhysRevE.102.032105, 8825523, PhysRevLett.130.107101},
as well as for discrete systems~\cite{PhysRevLett.126.010601, Dechant_2022, PhysRevResearch.5.013017, PhysRevX.13.011013}, Markovian open quantum systems~\cite{PhysRevLett.126.010601, PhysRevX.13.011013}, and nonlinear systems~\cite{PhysRevResearch.5.013017, PhysRevResearch.7.033011}.
More recently, optimal transport theory has been applied to information thermodynamics~\cite{PhysRevLett.109.180602, PhysRevX.4.031015, Parrondo:2015aa,Rosinberg_2016},
revealing fundamental bounds on the thermodynamic cost of performing information erasure~\cite{PhysRevLett.125.100602, PhysRevE.102.032105}, measurement and feedback~\cite{PhysRevResearch.3.043093, 9426929, PhysRevResearch.6.013023, PhysRevResearch.6.033239, td2s-819q, PhysRevResearch.7.013329, PhysRevResearch.7.023159} in finite time.

\begin{figure*}
    \centering
    \includegraphics[width=0.84\linewidth]{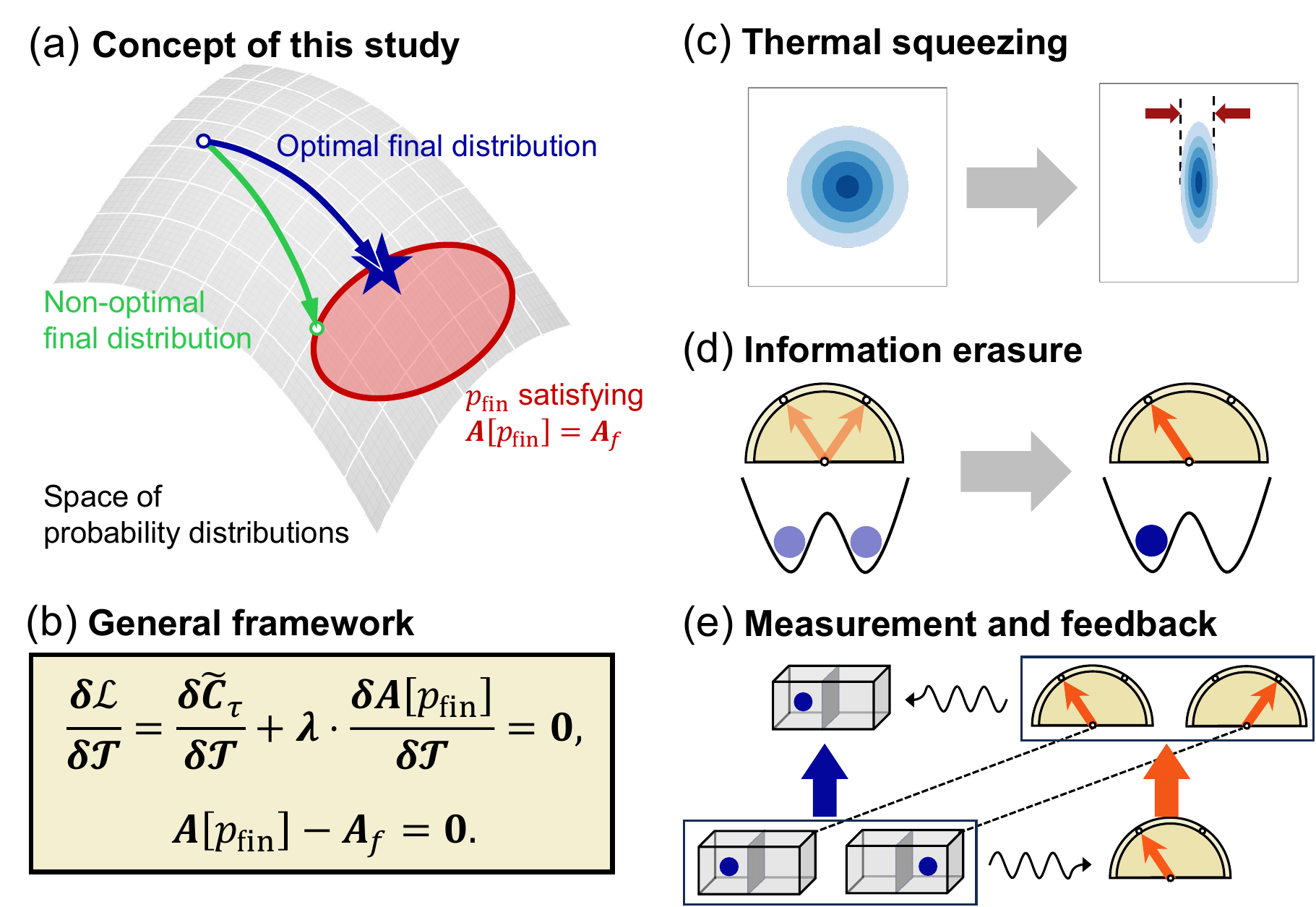}
    \caption{
    Summary of the main results and their applications.
    (a) Schematic illustration indicating that further optimization is possible beyond conventional optimal transport.
    Even if optimal transport protocols are used (blue and green arrows),
    choosing a non-optimal final distribution (green dot) yields a higher thermodynamic cost.
    Unlike the conventional setup that fixes the final distribution,
    we consider the freedom to choose a final distribution from the admissible set determined by the intended thermodynamic task.
    The boundary of the red region represents the set of admissible final distributions associated with each thermodynamic task,
    from which the optimal final distribution (blue star) is identified.
    (b) General framework that further optimizes the thermodynamic cost (Sec.~\ref{sec:general_framework}).
    We encode thermodynamic tasks as constraints on the final distributions, such that
    a quantity depending on the final distributions $\bmA[p_{\fin}]$ matches a desired value $\bmA_f$.
    The Lagrange multiplier method is employed to incorporate these constraints into the optimization problem.
    Solving the resulting variational equations yields the optimal transport map $\bmmcT^{\opt}$ that transports the initial probability distribution to the optimal final distribution.
    Applications of our framework (Sec.~\ref{sec:applications}): (c) thermal squeezing, (d) information erasure, and (e) measurement and feedback.
    }
    \label{fig:concept}
\end{figure*}

\begin{table*}[htbp]
    \centering
    \caption{Applications of our framework (see Sec.~\ref{sec:applications}).
    We show the thermodynamic cost $C_\tau$ and the constraints on the final distributions $\bmA[p_{\fin}]$ to perform given thermodynamic tasks.
    }
    \renewcommand{\arraystretch}{2.5}
    \begin{tabular}{l|l|l}
        \textbf{Thermodynamic tasks} & \textbf{Thermodynamic cost} $C_\tau$ & \textbf{Constraints at $t = \tau$ associated with the task:} $\bm{A}[p_{\fin}]$\\
        \hline \hline
        Particle transport & Entropy production $\Sigma_\tau$ & Expectation value of position: $\displaystyle \expval{\bm{r}}_{\fin}$\\
        \hline
        Thermal squeezing & $\Sigma_\tau$ & Covariance matrix of position: $\displaystyle \expval{(\bm{r} - \expval{\bm{r}}_{\fin})(\bm{r} - \expval{\bm{r}}_{\fin})^{\top}}_{\fin}$\\
        \hline
        Information erasure~\cite{PhysRevLett.125.100602, PhysRevE.102.032105} & $\Sigma_\tau$, Work $W_\tau$ & Cumulative distribution function at $r = 0$: $\int_{-\infty}^0 \dd r\, p_{\fin}(r)$\\
        \hline
        \makecell[l]{Work in systems with\\ quadratic potentials~\cite{8825523}} & $W_\tau$ & None\\
        \hline
        Control of free energy & $\Sigma_\tau$ & KL divergence: $D(p_{\fin} || q)$\\
        \hline
        Measurement and feedback & Partial entropy production $\Sigma_\tau^X, \Sigma_\tau^Y$ & Mutual information: $I_\fin$\\
        \hline
    \end{tabular}
    \label{table:fixed_quantities_applications}
\end{table*}

However, it is worth emphasizing that the straightforward application of optimal transport theory to thermodynamic tasks leaves room for further minimization of the thermodynamic cost.
For example, when we consider the task of thermal squeezing, which aims to reduce the variance along a certain direction in the phase space, there are infinitely many distributions that share the same variance.
Therefore, there remains room for further optimization of final distributions under the condition that the desired task is achieved (see Fig.~\ref{fig:concept}(a)).

In this paper, we develop a unified framework for finite-time optimization of the thermodynamic cost for various tasks in overdamped Langevin systems.
We take into account the freedom of the final distributions when performing thermodynamic tasks, and identify the optimal final distribution.
Our approach encodes thermodynamic tasks as constraints on the final distribution within the method of Lagrange multipliers,
thereby deriving variational equations that identify the optimal final distribution (see Fig.~\ref{fig:concept} (b)).
We can handle any kind of thermodynamic task as long as it can be expressed as equality constraints on functionals of the final distribution.
This includes the expectation values and variances of physical quantities of the final distributions.
We can also treat mutual information in a similar manner, which plays a crucial role in information processing.

Our framework applies broadly to thermodynamic tasks (see Fig.~\ref{fig:concept} and Table~\ref{table:fixed_quantities_applications}),
enabling us to obtain the truly optimal value of the thermodynamic cost and the protocol that achieves it.
We demonstrate our framework by optimizing the thermodynamic cost of particle transport, thermal squeezing, information erasure, work in quadratic potentials, control of free energy, measurement and feedback.
The framework also encompasses earlier studies, such as optimization of thermodynamic costs for information erasure~\cite{PhysRevLett.125.100602, PhysRevE.102.032105} and optimization of work in systems with quadratic potentials~\cite{8825523}.
The present framework extends the optimization of thermodynamic costs in systems far from equilibrium, beyond the slow-driving or linear response regime~\cite{PhysRevLett.99.100602, PhysRevLett.108.190602}.

The structure of this paper is as follows.
In Sec.~\ref{sec:notation}, we present notation, along with the thermodynamic speed limit based on optimal transport theory for overdamped Langevin systems.
In Sec.~\ref{sec:general_framework}, we develop the general framework for optimizing the thermodynamic cost required to perform thermodynamic tasks in finite time.
In Sec.~\ref{sec:applications}, we apply our framework to the tasks summarized in Table~\ref{table:fixed_quantities_applications}, showing that optimization of the thermodynamic cost required for such tasks can be treated.
In Sec.~\ref{sec:conclusion_and_discussion}, we conclude the paper with a summary and outlook.
In the Appendices, we provide details of the derivations of our results.

\section{Notation}\label{sec:notation}
We consider an overdamped Langevin system in $d$-dimensional Euclidean space in contact with a heat bath at temperature $T(= 1/\beta)$,
with mobility $\mu$ and diffusion coefficient $D = \mu T$.
Throughout this paper, we set the Boltzmann constant to $k_{\mathrm{B}} = 1$.
The system is driven by a conservative force $-\nabla V_t(\bmr)$ with $V_t (\bmr)$ being a time-dependent potential, and a nonconservative force $\bmF^{\mathrm{nc}}_t(\bmr)$.
The time evolution of the probability distribution $p_t(\bmr)$ is described by the Fokker--Planck equation~\cite{Kampen_Stochastic_Processes, Risken_Fokker_Planck_Equation}
\begin{equation}
    \partial_t p_t(\bmr) = - \nabla \cdot (\bmv_t(\bmr) p_t(\bmr)),
    \label{eq:fokker_planck}
\end{equation}
where $\bmv_t(\bmr)$ denotes the mean local velocity
\begin{equation}
    \bmv_t(\bmr) \coloneqq \mu \left(-\nabla V_t(\bmr) + \bmF^{\mathrm{nc}}_t(\bmr)\right) - D \nabla \ln p_t(\bmr).
    \label{eq:mean_local_velocity}
\end{equation}
We assume throughout the paper that the conservative force $- \nabla V_t(\bmr)$ and the nonconservative force $\bmF^{\mathrm{nc}}_t(\bmr)$ are fully controllable.

The entropy production $\Sigma_\tau$ is given by~\cite{Seifert_2012, Sekimoto_Stochastic_Energetics}
\begin{equation}
    \Sigma_\tau \coloneqq \frac{1}{D}\int_0^\tau \dd t \int \dd \bmr\, \|\bmv_t(\bmr)\|^2 p_t(\bmr) \geq 0,
    \label{eq:entropy_production}
\end{equation}
where $\|\cdot\|$ denotes the Euclidean norm.
The equality $\Sigma_\tau = 0$ holds if $\bmv_t(\bmr) = \bm{0}$ for all $t$, which is possible only in the quasistatic limit $\tau \to \infty$.
The work done on the system $W_\tau$ is given by~\cite{Sekimoto_Stochastic_Energetics, Seifert_2012}
\begin{equation}
    W_\tau \coloneqq \int_0^\tau \dd t \int \dd \bmr\, \left(\pdv{V_t(\bmr)}{t} + \bmF^{\mathrm{nc}}_t(\bmr) \cdot \bmv_t(\bmr)\right) p_t(\bmr).
\end{equation}
By using the average energy of the system $E_t \coloneqq \int \dd \bmr\, V_t(\bmr) p_t(\bmr)$ and the Shannon entropy $S_t \coloneqq -\int \dd \bmr\, p_t(\bmr) \ln p_t(\bmr)$, we define the nonequilibrium free energy as $\mcF_t \coloneqq E_t - T S_t$~\cite{Sekimoto_Stochastic_Energetics,Seifert_2012}.
We also define the equilibrium free energy $\mcF_t^{\mathrm{eq}} \coloneqq - \beta^{-1} \ln \int \dd \bmr\, e^{- \beta V_t(\bmr)}$.
The nonequilibrium free energy is expressed as $\mcF_t = \beta^{-1}D(p_t || p_t^{\mathrm{eq}}) + \mcF_t^{\mathrm{eq}}$, where $p_t^{\mathrm{eq}}(\bmr) \coloneqq e^{\beta(\mcF_t^{\mathrm{eq}} - V_t(\bmr))}$ and $D(p || q) \coloneqq \int \dd \bmr\, p(\bmr) \ln (p(\bmr)/q(\bmr))$ is the Kullback--Leibler divergence.
We note that the work can be expressed as $W_\tau = \Delta \mcF + T \Sigma_\tau$, where $\Delta \mcF \coloneqq \mcF_\tau - \mcF_0$ is the change in the nonequilibrium free energy.

We fix the initial distribution $p_{\ini}$, the final distribution $p_{\fin}$, and the operation time $\tau$.
The Benamou--Brenier formula~\cite{Benamou:2000aa} implies that the minimum entropy production is given by the $L^2$-Wasserstein distance from optimal transport theory~\cite{Aurell:2012aa, PhysRevResearch.3.043093}:
\begin{equation}
    \min_{\substack{\{p_t, \bmv_t\}_{0\leq t \leq \tau} \\ \text{s.t.}\, p_{0}=p_{\ini},\, p_{\tau}=p_{\fin}}} \Sigma_\tau = \frac{\mcW_2(p_{\ini}, p_{\fin})^2}{D \tau}.
    \label{eq:EP_speed_limit}
\end{equation}
Here, the minimum is taken over all protocols $\{p_t, \bmv_t\}_{0\leq t \leq \tau}$ that satisfy Eq.~\eqref{eq:fokker_planck} with boundary conditions $p_{t=0}=p_{\ini}$ and $p_{t=\tau}=p_{\fin}$.
The squared $L^2$-Wasserstein distance between $p_{\ini}$ and $p_{\fin}$ is defined as~\cite{Villani_optimal_transport, Santambrogio_Optimal_Transport}
\begin{eqnarray}
    &&\mcW_2(p_{\ini}, p_{\fin})^2 \coloneqq\nonumber\\
    &&\min_{\substack{\bmmcT\, \text{s.t.}\, \bmmcT_{\sharp} p_{\ini} = p_{\fin}}} \int \dd \bmr\, \|\bmmcT(\bmr) - \bmr\|^2 p_{\ini}(\bmr),
    \label{eq:Wasserstein_Monge}
\end{eqnarray}
where, following the convention in mathematics, $\bmmcT_{\sharp} p_{\ini} = p_{\fin}$ means that the minimum in Eq.~\eqref{eq:Wasserstein_Monge} is taken over transport maps $\bmmcT$ pushing $p_{\ini}$ forward to $p_{\fin}$.
That is, $\bmmcT_{\sharp} p_{\ini} = p_{\fin}$ implies that $\int \dd\bmr\, f\!\left(\bmmcT(\bmr)\right) p_{\ini}(\bmr) = \int \dd \bmr\, f(\bmr) p_{\fin} (\bmr)$ for any continuous function $f$,
which, when $\bmmcT$ is sufficiently smooth and has a nonsingular Jacobian $J_{\bmmcT}(\bmr) \coloneqq \pdv{\bmmcT(\bm{r})}{\bmr}$, can be written as
\begin{equation}
    p_{\ini}(\bmr) = p_{\fin}\left(\bmmcT(\bmr)\right) \det J_{\bmmcT}(\bmr).
    \label{eq:change_variable}
\end{equation}

\section{General framework}\label{sec:general_framework}
In this section, we present our general framework.
The problem we address is as follows: given an initial distribution $p_{\ini}$, we optimize the thermodynamic cost $C_\tau$ required to perform various thermodynamic tasks while the system evolves according to Eq.~\eqref{eq:fokker_planck} over the finite fixed time $\tau$.
In what follows, depending on the thermodynamic task, we take as the thermodynamic cost the entropy production $\Sigma_\tau$, the work $W_\tau$, or the partial entropy production $\Sigma_\tau^i$ $(i = X, Y)$ in bipartite systems, as listed in Table~\ref{table:fixed_quantities_applications}.

\subsection{Setup of optimization}\label{subsec:setup_of_optimization}
First, we define a quantity $\bm{A}[p_{\fin}]$ dependent on the final distribution $p_{\fin}$,
which includes various examples as listed in Table~\ref{table:fixed_quantities_applications}.
Then, we encode thermodynamic tasks as constraints on $\bmA[p_\fin]$ such that it matches a desired value $\bmA_f$.
The crucial point is that there are (in general, infinitely) many distributions that satisfy this condition.
Therefore, in addition to optimizing the protocol that achieves the minimum in Eq.~\eqref{eq:EP_speed_limit}, the final distribution $p_{\fin}$ can be further optimized.

We consider the thermodynamic cost $C_\tau$ required for achieving the above task.
In general, we suppose that $C_\tau$ depends on the time evolution of the distribution and the mean local velocity, which we explicitly write as $C_\tau \left[\{ p_t, \bmv_t \}_{0 \leq t \leq \tau}\right]$.
Then, the optimization problem can be formulated as
\begin{equation}
    \min_{p_{\fin}\, \text{s.t.}\, \bmA[p_{\fin}] = \bmA_f} \min_{\substack{\{p_t, \bmv_t\}_{0\leq t \leq \tau} \\ \text{s.t.}\, p_{0} = p_{\ini},\, p_{\tau} = p_{\fin}}} C_\tau[\{p_t, \bmv_t\}_{0 \leq t \leq \tau}].
    \label{eq:optimization_problem}
\end{equation}

We now show that the optimization problem Eq.~\eqref{eq:optimization_problem} can be reduced to a more tractable expression.
We first consider the special case where $C_\tau=\Sigma_\tau$.
For a transport map $\bmmcT$, we define
\begin{eqnarray}
    \tilde{\Sigma}_\tau\left[\bmmcT\right] \coloneqq \frac{1}{D\tau} \int \dd \bmr\, \|\bmmcT(\bmr) - \bmr\|^2 p_{\ini}(\bmr).
    \label{eq:Sigma_tilde}
\end{eqnarray}
From Eqs.~\eqref{eq:EP_speed_limit} and~\eqref{eq:Wasserstein_Monge}, we have
\begin{eqnarray}
    && \min_{\substack{\{p_t, \bmv_t\}_{0\leq t \leq \tau} \\ \text{s.t.}\, p_{0} = p_{\ini},\, p_{\tau} = p_{\fin}}} \Sigma_\tau[\{p_t, \bmv_t\}_{0 \leq t \leq \tau}] \nonumber\\
    &&=\min_{\bmmcT\, \text{s.t.}\, \bmmcT_{\sharp} p_{\ini} = p_{\fin}} \tilde{\Sigma}_\tau[\bmmcT].
    \label{eq:identity_Sigma_tilde_Sigma}
\end{eqnarray}
In addition, since $p_{\fin}$ is a functional of $\bmmcT$ via Eq.~\eqref{eq:change_variable}, $\bmA[p_{\fin}]$ can be written as $\bmA[\bmmcT]$.
Therefore, our optimization problem Eq.~\eqref{eq:optimization_problem} reduces to
\begin{eqnarray}
    &&\min_{\substack{p_{\fin} \text{ s.t. } \bmA[p_{\fin}] = \bmA_f}} \min_{\bmmcT \text{ s.t. } \bmmcT_{\sharp} p_{\ini} = p_{\fin}} \tilde{\Sigma}_\tau[\bmmcT]\nonumber\\
    &&=\min_{\bmmcT \text{ s.t. } \bmA[\bmmcT] = \bmA_f} \tilde{\Sigma}_\tau[\bmmcT].
\end{eqnarray}
A similar argument applies to other $C_\tau$ listed in Table~\ref{table:fixed_quantities_applications},
by defining $\tilde{C}_\tau[\bmmcT]$ appropriately (see Secs.~\ref{subsubsec:variation_work} and~\ref{subsec:measurement_feedback} for details).

In summary, we consider the optimization problem equivalent to Eq.~\eqref{eq:optimization_problem}, which takes the simple form
\begin{equation}
    \min_{\bmmcT\, \text{s.t.}\, \bmA[\bmmcT] = \bmA_f} \tilde{C}_\tau[\bmmcT].
    \label{eq:optimization_problem_transport_map}
\end{equation}

\subsection{Main formula}
We adopt a Lagrange multiplier approach that takes the transport map as the optimization variable.
The Lagrangian $\mcL$ contains the transport map $\bmmcT$ and the Lagrange multipliers $\bmlambda$ as variables, and is given by
\begin{equation}
    \mathcal{L}[\bmmcT, \bmlambda] = \tilde{C}_\tau[\bmmcT] + \bmlambda \cdot \left(\bmA[\bmmcT] - \bmA_f\right).
    \label{eq:Lagrangian}
\end{equation}
The first-order optimality conditions read
\begin{equation}
    \frac{\delta \tilde{C}_\tau[\bmmcT]}{\delta \bmmcT} + \bmlambda \cdot \frac{\delta \bmA[\bmmcT]}{\delta \bmmcT} = 0,
    \label{eq:Lagrangian_variation_transport_map}
\end{equation}
\begin{equation}
    \bmA[\bmmcT] - \bmA_f = 0.
    \label{eq:constraint}
\end{equation}
Solving Eqs.~\eqref{eq:Lagrangian_variation_transport_map} and~\eqref{eq:constraint} yields solutions $(\bmmcT^{\opt}, \bmlambda^{\opt})$,
from which the optimal value of the thermodynamic cost $C_\tau^{\opt} = \tilde{C}_\tau[\bmmcT^{\opt}]$, the optimal final distribution $p_{\fin}^{\opt}$, and the optimal protocol $\{p_t^{\opt}, \bmv^{\opt}_t\}_{0\leq t \leq \tau}$ follow.
The explicit form of the optimal mean local velocity is given by~\cite{Benamou:2000aa, Aurell:2012aa}
\begin{equation}
    \bmv_t^{\opt}\left(\left(1 - \frac{t}{\tau}\right)\bmr + \frac{t}{\tau}\bmmcT^\opt(\bmr)\right) = \frac{\bmmcT^\opt(\bmr) - \bmr}{\tau}.
\end{equation}
These constitute the general framework shown in Fig.~\ref{fig:concept}(b).
We note that a similar approach has also been adopted previously in the setting of information erasure~\cite{PhysRevLett.125.100602, PhysRevE.102.032105}.

Throughout, we use the Einstein summation convention.
We note that the details of the derivations are provided in Appendix~\ref{ap_sec:derivation_main_results}.

\subsection{Variation of thermodynamic costs \texorpdfstring{$C_\tau$}{C}}~\label{subsec:variation_thermodynamic_costs}
As mentioned before, the thermodynamic cost $C_\tau$ we consider is the entropy production $\Sigma_\tau$, the work $W_\tau$, or the partial entropy production $\Sigma_\tau^i\ (i = X, Y)$.
In this subsection, we present the variational formulas for $\Sigma_\tau$ and $W_\tau$,
while $\Sigma_\tau^i\ (i = X, Y)$ is discussed later in Sec.~\ref{subsec:measurement_feedback}.
What we mainly treat below is the first term of Eq.~\eqref{eq:Lagrangian_variation_transport_map}.

\subsubsection{\texorpdfstring{$C_\tau$}{C}: entropy production}
We first consider the entropy production.
From Eq.~\eqref{eq:Sigma_tilde}, the variation of $\tilde{\Sigma}_\tau[\bmmcT]$ with respect to the transport map $\bmmcT$ is given by
\begin{equation}
    \frac{\delta \tilde{\Sigma}_\tau[\bmmcT]}{\delta \bmmcT} = \frac{2}{D\tau} (\bmmcT(\bmr) - \bmr) p_{\ini}(\bmr),
    \label{eq:variation_EP}
\end{equation}
which is linear in $\bmmcT$.

\subsubsection{\texorpdfstring{$C_\tau$}{C}: work}\label{subsubsec:variation_work}
We next consider the work.
When optimizing the work in finite time, we impose the boundary condition that the potentials $V_t$ at $t = 0$ and $t = \tau$ are fixed to $V_\ini$ and $V_\fin$, respectively~\cite{PhysRevLett.98.108301}.
Otherwise, even in the quasistatic limit, $W = \mcF^{\mathrm{eq}}_{\fin} - \mcF_\ini$ could be made arbitrarily small by varying $V_0$ or $V_\tau$.
Here we fix $V_0$ and $V_\tau$ to eliminate this triviality.

Using $\tilde{\Sigma}_\tau[\bmmcT]$ in Eq.~\eqref{eq:Sigma_tilde}, we define $\tilde{W}_\tau[\bmmcT] \coloneqq \Delta \mcF[\bmmcT] + \beta^{-1} \tilde{\Sigma}_\tau[\bmmcT]$.
We note that since $\Delta \mcF[\bmmcT]$ depends only on $p_\ini$, $p_\fin$, $V_\ini$ and $V_\fin$, the optimization problem~\eqref{eq:optimization_problem} with $C_\tau = W_\tau$ is equivalent to Eq.~\eqref{eq:optimization_problem_transport_map} with $\tilde{C}_\tau[\bmmcT] = \tilde{W}_\tau[\bmmcT]$.
The variation of $\tilde{\Sigma}_{\tau}[\bmmcT]$ was obtained in Eq.~\eqref{eq:variation_EP}, so it remains to vary $\mcF_\fin[\bmmcT]$, which is dependent on $\bmmcT$ via $p_\fin$.
Defining $p_\fin^{\mathrm{eq}}(\bmr) \coloneqq e^{- \beta V_\fin(\bmr)}/\int \dd \bmr' e^{- \beta V_\fin(\bmr')}$, the dependence of $\mcF_\fin[\bmmcT]$ on the transport map is given by
\begin{eqnarray}
    \mathcal{F}_{\fin}[\bmmcT] = \beta^{-1} \int \dd \bm{r}\, p_{\ini}(\bm{r}) \ln \frac{p_{\ini}(\bm{r})}{p_{\fin}^{\mathrm{eq}}(\bm{\mathcal{T}}(\bm{r})) \det J_{\bm{\mathcal{T}}}(\bm{r})}+ \mathcal{F}_{\fin}^{\mathrm{eq}}.\nonumber\\
    \label{eq:free_energy_tau}
\end{eqnarray}
Thus, we have
\begin{eqnarray}
    \frac{\delta \mcF_\fin[\bmmcT]}{\delta \mcT_i} &=& \beta^{-1} \left[- p_{\ini}(\bm{r})\left. \partial_{r_i'} \ln p_{\fin}^{\mathrm{eq}}(\bm{r}')\right|_{\bm{r}' = \bm{\mathcal{T}}(\bm{r})}\right.\nonumber\\
    &\phantom{=}&+ \left. \partial_{r_j} \left\{p_{\ini}(\bm{r}) \tilde{J}_{ji}(\bmr) \right\} \right].
    \label{eq:free_energy_variation}
\end{eqnarray}
See Appendix~\ref{ap_subsec:variation_free_energy} for the details of the derivation.
Here, $\tilde{J}_{ji}(\bmr) \coloneqq (J_{\bm{\mcT}}(\bmr)^{-1})_{ji} = (J_{\bm{\mcT}^{-1}}(\bm{\mcT}(\bmr)))_{ji} = \partial_{r_i'} \mcT_j^{-1}(\bmr')|_{\bmr' = \bm{\mcT}(\bmr)}$.
Therefore, the variation of $\tilde{W}_\tau[\bmmcT]$ is given by
\begin{eqnarray}
    \frac{\delta \tilde{W}_\tau[\bmmcT]}{\delta \mcT_i} &=& \beta^{-1} \left[- p_{\ini}(\bm{r})\left. \partial_{r_i'} \ln p_{\tau}^{\mathrm{eq}}(\bm{r}')\right|_{\bm{r}' = \bm{\mathcal{T}}(\bm{r})}\right.\nonumber\\
    &\phantom{=}&+ \left. \partial_{r_j} \left\{p_{\ini}(\bm{r}) \tilde{J}_{ji}(\bmr) \right\}\right.\nonumber\\
    &\phantom{=}&\left.+ \frac{2}{D\tau} (\mcT_i(\bmr) - r_i) p_{\ini}(\bmr) \right].
    \label{eq:variation_work}
\end{eqnarray}
Although the potentials at $t = 0$ and $t = \tau$ are fixed, the potential shape may change discontinuously at those times.
Indeed, it is known that quenches at $t = 0$ and $t = \tau$ are required in finite-time thermodynamic optimization~\cite{PhysRevLett.98.108301}.

\subsection{Variation of \texorpdfstring{$\bmA[p_{\fin}]$}{A}} \label{subsec:variation_fixed_quantities}
We next derive the variational formula for $\bmA[p_{\fin}] = \bmA[\bmmcT]$ with respect to the transport map.
Combining it with the variation of $C_\tau$ from the previous subsection yields the optimal final distributions $p_{\fin}^{\opt}$ via Eqs.~\eqref{eq:Lagrangian_variation_transport_map} and~\eqref{eq:constraint}.

In general, a quantity $\bmA[\bmmcT]$ can be written as a functional of the final distribution $p_{\fin}[\bmmcT]$ of the form $\int \dd \bmr\, \bmg(p_{\fin}[\bmmcT(\bmr)], \bmr) = \int \dd \bmr\, \det J_{\bmmcT}(\bmr) \bmg \left(\frac{p_{\ini}(\bmr)}{\det J_{\bmmcT}(\bmr)}, \bmmcT(\bmr)\right)$.
For example, if the thermodynamic task is to make the expectation value of a physical observable $\bma(\bmr)$ at the final time $\expval{\bma}_{\fin}$ equal to a desired value,
then $\bmg \left(p_{\fin}[\bmmcT(\bmr)], \bmr\right) = \bma(\bmr)\, p_{\fin}[\bmmcT(\bmr)]$.
Also, the Shannon entropy is described by taking $g\left(p_{\fin}[\bmmcT(\bmr)], \bmr\right) = - p_{\fin}[\bmmcT(\bmr)] \ln p_{\fin}[\bmmcT(\bmr)]$.
For measurement and feedback tasks, we fix the mutual information of the final distributions.
Although mutual information does not take the above form because marginalization of the final probability distribution is involved,
it can still be treated in a similar manner, as we discuss in Sec.~\ref{sec:applications}.

For $\bmA[\bmmcT] = \int \dd \bmr\, \bmg \left(p_{\fin}[\bmmcT(\bmr)], \bmr\right)$, the variation with respect to $\mcT_i$ is given by
\begin{equation}
    \frac{\delta \bmA[\bmmcT]}{\delta \mcT_i} = \int \dd \bm{r}'\, \left.\pdv{\bmg(X, \bmr')}{X}\right|_{X = p_{\fin}(\bmr')} \frac{\delta p_{\fin}(\bmr')}{\delta \mcT_i(\bmr)}.
\end{equation}
By further using the expression for $\frac{\delta p_\fin[\bmmcT]}{\delta \mcT_i}$ obtained in Appendix~\ref{ap_subsec:variation_functional_final_distributions}, we have
\begin{equation}
    \frac{\delta \bmA[\bmmcT]}{\delta \mcT_i} = \partial_{r_j} \left\{\hat{\bmg} (\bm{\mathcal{T}}(\bmr)) \right\} \tilde{J}_{ji}(\bm{r}) p_{\ini}(\bmr),
    \label{eq:variation_functional_final_distributions}
\end{equation}
where $\hat{\bmg}(\bmmcT(\bm{r})) = \left.\pdv{\bmg(X, \bmmcT(\bmr))}{X}\right|_{X = p_{\fin}(\bmmcT(\bm{r}))}$.
Equation~\eqref{eq:variation_functional_final_distributions} is the general form for the variation of a quantity $\bmA[\bmmcT]$ associated with the thermodynamic task.
In Sec.~\ref{sec:applications} we treat variations of specific quantities depending on the intended task.

An important choice of the quantity $\bmA[\bmmcT]$ is the expectation value of a physical observable at the final time $\bmA[\bmmcT] = \expval{\bma}_{\fin}[\bmmcT] = \int \dd \bmr\, \bma(\bmr) p_{\fin}[\bmmcT(\bmr)]$.
In this case we obtain simpler variational formulas.
Since Eq.~\eqref{eq:change_variable} implies
\begin{equation}
    \expval{\bma}_{\fin}[\bmmcT] = \int \dd \bmr\, \bma(\bmmcT(\bmr)) p_{\ini}(\bmr),
\end{equation}
we obtain by direct calculation
\begin{equation}
    \frac{\delta \expval{\bma}_{\fin}[\bmmcT]}{\delta \bmmcT} = \left. \partial_{\bmr'} \bma(\bmr')\right|_{\bmr' = \bmmcT(\bmr)} p_{\ini}(\bmr).
    \label{eq:variation_expectation_value_physical_quantities}
\end{equation}
We note that Eq.~\eqref{eq:variation_expectation_value_physical_quantities} is consistent with Eq.~\eqref{eq:variation_functional_final_distributions} when $\bmg(p_{\fin}[\bmmcT(\bmr)], \bmr) = \bma(\bmr) p_{\fin}[\bmmcT(\bmr)]$ (see Appendix~\ref{ap_subsec:variation_functional_final_distributions} for details).

While our framework provides a necessary condition for global optimality in general,
it guarantees global optimality if a feasible stationary solution $(\bmmcT^*, \bmlambda^*)$ exists and the Lagrangian density $\mathscr{L}(\bmr, \bmmcT, \pdv{\bmmcT}{\bmr}, \bmlambda)$, defined via $\mathcal{L} = \int \dd \bmr\, \mathscr{L}$, is jointly convex in $(\bmmcT, \pdv{\bmmcT}{\bmr})$ for all $\bmr$ with $\bmlambda = \bmlambda^*$~\cite{Clarke:2013aa}.
One can verify that the density for the entropy production is convex in $\bmmcT$, as is the Lagrangian density for the partial entropy production introduced in Sec.~\ref{subsec:measurement_feedback}.
However, the convexity of the density for the work and the density for $\bmA[\bmmcT]$ depends on the specific forms of the potential and constraints.
In Appendix~\ref{ap_sec:several_remarks}, we further comment on imposing inequality constraints, the differentiability of the transport map, and the numerical cost of solving the optimization problem.

\section{Applications} \label{sec:applications}
\begin{figure*}[tb]
    \centering
    \includegraphics[width=0.9\linewidth]{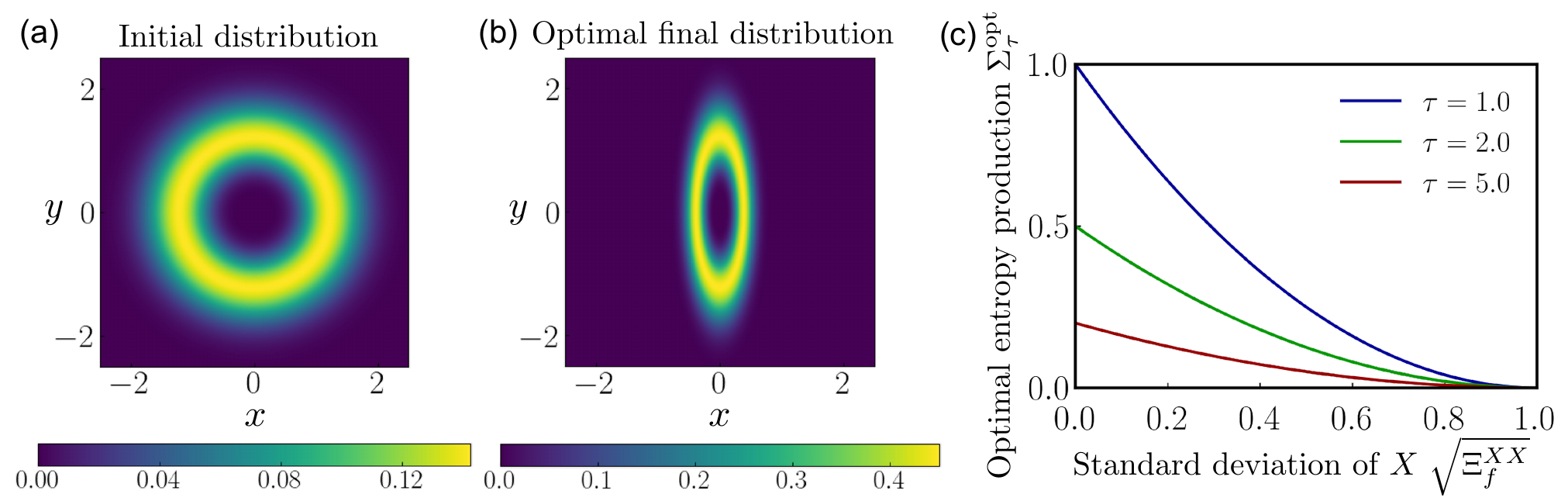}
    \caption{Non-Gaussian example of thermodynamically optimal thermal squeezing in finite time.
    (a) Initial distribution before performing thermal squeezing.
    We take as the initial distribution the intensity profile of a normalized Laguerre--Gaussian mode with radial index~0 and azimuthal index~3~\cite{Kogelnik:66}, which is a non-Gaussian distribution.
    (b) Final distribution after performing thermal squeezing with $\Xi_f^{XX} = 0.1$.
    (c) Dependence of the minimum entropy production $\Sigma_\tau^{\opt}$ on the standard deviation of $X$ in the final distributions $\sqrt{\Xi_f^{XX}}$.
    This plot illustrates the trade-off relation that the smaller the variance of the final distributions, the larger the minimal entropy production required.
    The required minimum entropy production also decreases as the operation time $\tau$ increases.
    The diffusion coefficient is set to $D = 1$.}
    \label{fig:thermal_squeezing}
\end{figure*}
Using our framework developed in Sec.~\ref{sec:general_framework},
we demonstrate its applicability to the thermodynamic tasks listed in Table~\ref{table:fixed_quantities_applications}.
Throughout, the expectation value of a physical observable $\bma (\bmr)$ with respect to $p_{\ini}$ is denoted by $\expval{\bma(\bmr)}_{\ini}$.

\subsection{Particle transport}\label{subsec:particle_transport}

The simplest application of our framework is particle transport, that is, the task of transporting a particle from its initial position to a target position.
We take the entropy production $\Sigma_\tau$ as the thermodynamic cost $C_\tau$.
We focus on transport problems in which only the expectation value of position is relevant.
Accordingly, we impose on the final distribution the constraint $\bmA[p_{\fin}] = \expval{\bmr}_{\fin} = \bmmu_f$, where $\bmmu_f$ is the target position.
From Eqs.~\eqref{eq:variation_EP} and~\eqref{eq:variation_expectation_value_physical_quantities}, the equations corresponding to Eqs.~\eqref{eq:Lagrangian_variation_transport_map} and~\eqref{eq:constraint} are given by
\begin{equation}
    \frac{2 p_{\ini}(\bmr)}{D \tau} \left(\bm{\mcT}(\bmr) - \bmr\right) + \bmlambda p_{\ini}(\bmr) = 0,
    \label{eq:fixed_position_variation}
\end{equation}
\begin{equation}
    \int \dd \bmr\, \bm{\mcT}(\bmr) p_{\ini}(\bmr) - \bmmu_f = 0,
    \label{eq:fixed_position_constraints}
\end{equation}
which can be solved straightforwardly to yield
\begin{equation}
    \bm{\mcT}^{\opt}(\bmr) = \bmr + \bmmu_f - \bmmu_\ini,
\end{equation}
\begin{equation}
    p_{\fin}^{\opt}(\bmr) = p_{\ini}(\bmr - \bmmu_f + \bmmu_\ini),
\end{equation}
where we define $\bmmu_\ini \coloneqq \expval{\bmr}_{\ini}$.

\subsection{Thermal squeezing}\label{subsec:thermal_squeezing}
Thermal squeezing is the reduction of fluctuations along a certain direction in phase space below that of a thermal state.
Thermally squeezed states, when used as information carriers, enable information erasure beyond the Landauer bound~\cite{PhysRevLett.122.040602}, and when used as a heat bath, allow the design of heat engines beyond the Carnot bound~\cite{PhysRevLett.112.030602, PhysRevX.7.031044}.
Such thermal squeezing has been realized experimentally in mechanical oscillators~\cite{PhysRevLett.67.699} and in nanoparticles~\cite{PhysRevLett.117.273601, doi:10.1126/science.ady4652}.

In Sec.~\ref{subsec:particle_transport} we constrained only the expectation value of position $\bmr$ with respect to $p_{\fin}$.
Here, by additionally constraining the covariance of final distributions, we apply our framework to thermal squeezing.
We take the entropy production $\Sigma_\tau$ as the thermodynamic cost $C_\tau$, and impose the constraints $\expval{\bmr}_{\fin} = \bmmu_f$ and $\expval{\left(\bmr - \expval{\bmr}_{\fin}\right) \left(\bmr - \expval{\bmr}_{\fin}\right)^\top}_{\fin} = \Xi_f$ on the expectation and covariance of position, where $\cdot^\top$ represents the transpose.

Here, we derive the optimal transport map and the minimal entropy production for thermal squeezing.
The arguments of the Lagrangian Eq.~\eqref{eq:Lagrangian} are the transport map $\bmmcT(\bmr)$ and the Lagrange multipliers $\bmnu$ and $\Lambda$ corresponding to constraining the expectation value and the covariance of position at the final time.
Since the covariance matrix is a positive-definite symmetric matrix, $\Lambda$ is taken to be symmetric.
The term corresponding to constraining the covariance matrix is $\Tr[\Lambda\left(\expval{\left(\bmr - \expval{\bmr}_{\fin}\right) \left(\bmr - \expval{\bmr}_{\fin}\right)^\top}_{\fin} - \Xi_f\right)]$.
Using Eqs.~\eqref{eq:variation_EP} and~\eqref{eq:variation_expectation_value_physical_quantities}, the equations corresponding to Eqs.~\eqref{eq:Lagrangian_variation_transport_map} and~\eqref{eq:constraint} are given by
\begin{align}
    &\frac{2 p_{\ini}(\bm{r})}{D \tau} \left(\bm{\mathcal{T}}(\bm{r}) - \bm{r}\right) + p_{\ini}(\bm{r})\bmnu\nonumber\\
    &+ 2 p_{\ini}(\bm{r}) \Lambda (\bm{\mathcal{T}}(\bm{r}) - \bm{\mu}_f) = 0,
    \label{eq:fixed_position_covariance_variation}
\end{align}
\begin{equation}
    \int \dd \bmr\, p_{\ini}(\bmr) \bm{\mathcal{T}}(\bmr) - \bmmu_f = 0,
    \label{eq:fixed_position_exp_constraints}
\end{equation}
\begin{equation}
    \int \dd \bmr\, p_{\ini}(\bmr) \bm{\mathcal{T}}(\bmr) \bm{\mathcal{T}}(\bmr)^\top - \bmmu_f \bmmu_f^\top - \Xi_f = 0.
    \label{eq:fixed_position_cov_constraints}
\end{equation}
Solving these equations yields the optimal transport map
\begin{equation}
    \bm{\mathcal{T}}^{\opt}(\bmr) = \Xi_\ini^{-\frac{1}{2}} \left(\Xi_\ini^{\frac{1}{2}} \Xi_f \Xi_\ini^{\frac{1}{2}}\right)^{\frac{1}{2}} \Xi_\ini^{-\frac{1}{2}} \left(\bmr - \bmmu_\ini\right) + \bmmu_f,
    \label{eq:transportmap_covariance_fixed}
\end{equation}
where $\bmmu_\ini \coloneqq \expval{\bmr}_{\ini}$ and $\Xi_\ini \coloneqq \expval{\left(\bmr - \expval{\bmr}_{\ini}\right) \left(\bmr - \expval{\bmr}_{\ini}\right)^\top}_{\ini}$.
Thus, the optimal transport map is affine in $\bmr$.
The optimal entropy production is given by
\begin{eqnarray}
    \Sigma_\tau^{\opt} &=& \frac{1}{D\tau} \left(\left\|\bmmu_f - \bmmu_\ini\right\|^2 \right.\nonumber\\
                       &\phantom{=}& \left.+ \Tr[\Xi_\ini + \Xi_f - 2 \left(\Xi_\ini^{\frac{1}{2}} \Xi_f \Xi_\ini^{\frac{1}{2}}\right)^\frac{1}{2}]\right).
    \label{eq:entropy_production_covariance_fixed}
\end{eqnarray}

As an example, we demonstrate thermodynamically optimal thermal squeezing for a non-Gaussian initial distribution in two dimensions (see Fig.~\ref{fig:thermal_squeezing}).
Figure~\ref{fig:thermal_squeezing}(c) shows the trade-off relation that the smaller the variance in $x$ of the final distributions, the larger the entropy production required for thermal squeezing.

We demonstrate that affine transport maps are optimal for particle transport and thermal squeezing even for non-Gaussian distributions, as illustrated in Fig.~\ref{fig:thermal_squeezing}.
We note that the optimal solutions in Eqs.~\eqref{eq:transportmap_covariance_fixed} and~\eqref{eq:entropy_production_covariance_fixed} follow directly from the results of Ref.~\cite{https://doi.org/10.1002/mana.19901470121}.
While it is sufficient to use quadratic potentials to implement optimal transport for Gaussian distributions~\cite{Aurell:2012aa},
that is not the case for non-Gaussian distributions, including optimal thermal squeezing for non-Gaussian distributions discussed here.

\subsection{Work in systems with quadratic potentials}\label{subsubsec:work_gaussian_processes}
The probability distributions of overdamped Langevin systems trapped in quadratic potentials remain Gaussian if the initial distributions are Gaussian.
Due to the analytical tractability of this setting, previous studies have derived the finite-time optimal control protocol~\cite{PhysRevLett.98.108301} and the efficiency of a Brownian heat engine at maximum power~\cite{Schmiedl_2008}.

Using our framework, we can analytically optimize the work in this setting and recover the results of earlier work~\cite{8825523}.
This corresponds to taking $C_\tau = W_\tau$ and imposing no constraint on the final distributions.
Given a Gaussian initial distribution, we minimize the work using a quadratic potential $V_t(\bmr) = \frac{1}{2} \left(\bmr - \bmp(t)\right)^\top Q(t) \left(\bmr - \bmp(t)\right)$,
where $\bmp(t)$ is the center of the potential and $Q(t)$ is its strength, which is a positive-definite symmetric matrix.
As in Sec.~\ref{subsubsec:variation_work}, we fix the potentials at $t = 0$ and $t = \tau$.
Since $p_t(\bmr)$ is Gaussian at all times, the optimal transport map is restricted to the form
\begin{equation}
    \bmmcT(\bmr) = \Xi_\ini^{- \frac{1}{2}} \left(\Xi_\ini^{\frac{1}{2}} \Xi_\fin \Xi_\ini^{\frac{1}{2}}\right)^{\frac{1}{2}} \Xi_\ini^{- \frac{1}{2}} (\bmr - \bmmu_\ini) + \bmmu_\fin,
\end{equation}
where $\bmmu_\ini, \bmmu_\fin, \Xi_\ini$, and $\Xi_\fin$ are the means and covariance matrices of the initial and final distributions, respectively.
Substituting this into the variational equation~\eqref{eq:variation_work} and solving for $\bmmu_\fin$ and $\Xi_\fin$ reproduces the results of Ref.~\cite{8825523}.

\subsection{Information erasure}\label{subsec:information_erasure}
We next consider information erasure as an application.
Together with measurement and feedback discussed later, information erasure is an elementary operation of information processing.
According to Landauer's principle~\cite{10.1063/1.881299}, erasing one bit of information stored in a symmetric double-well potential necessarily generates at least $k_\mathrm{B} T \ln 2$ of heat,
which has been verified experimentally~\cite{Berut:2012aa, PhysRevLett.129.270601, PhysRevLett.113.190601, PhysRevLett.126.170601, doi:10.1073/pnas.2301742120, PhysRevE.111.044114}.
The problem of optimizing the entropy production and the work required for finite-time information erasure has been addressed in earlier studies~\cite{PhysRevLett.125.100602,PhysRevE.102.032105}
by constraining the cumulative distribution function at $r = 0$, given as $\int_{-\infty}^{0}\dd r\, p_{\fin}(r) $, which represents the probability that the particle is in the ``0'' state.
In our framework, choosing the same thermodynamic cost $C_\tau$ and the same quantity $A[p_\fin]$ reproduces the results of Refs.~\cite{PhysRevLett.125.100602, PhysRevE.102.032105}.

On the other hand, by choosing other quantities as $\bmA[p_\fin]$, one can examine the erasure process from a different point of view.
We explore this possibility in Appendix~\ref{ap_sec:example_information_erasure} by taking the exponentially weighted average of the position as $\bmA[p_\fin]$.

\subsection{Control of free energy}\label{subsec:control_free_energy}
We next take the nonequilibrium free energy as the quantity to be constrained.
For a system in contact with a single heat bath, the fundamental bound on the extractable work is given by the change in nonequilibrium free energy, so preparing states with large free energy is important for designing work-storage devices.
From this viewpoint, optimization of free energy has also been studied in the context of free-energy harvesting~\cite{PhysRevResearch.6.013275, e27010091}.

In this case, the variation of the free energy with respect to the transport map is given by Eq.~\eqref{eq:free_energy_variation}.
Using our framework, one can determine the optimal transport map and the optimal final distribution that yields the desired change in free energy.
As a demonstration, we provide an example of optimal control of free energy in Appendix~\ref{ap_sec:example_free_energy_control}.

\subsection{Measurement and feedback}\label{subsec:measurement_feedback}
Besides information erasure, fundamental tasks in information processing include measurement and feedback.
In the setting of information thermodynamics, the second law of thermodynamics has been generalized to include mutual information~\cite{Parrondo:2015aa}.
Recent studies have extended these results to finite-time processes using optimal transport theory~\cite{PhysRevResearch.3.043093, 9426929, PhysRevResearch.6.013023, PhysRevResearch.6.033239, td2s-819q, PhysRevResearch.7.013329, PhysRevResearch.7.023159}.
Among these, studies on overdamped Langevin systems~\cite{9426929, PhysRevResearch.7.023159} fix the final distributions, leaving room for further optimization.

\subsubsection{Setup for measurement and feedback}\label{subsubsec:setup_information_thermodynamics}
We now introduce the setup of information thermodynamics.
We consider a bipartite system $XY$, with $X$ being the system of interest and $Y$ the memory.
The subsystems $X$ and $Y$ are each $d$-dimensional, and the total system decomposes as $\bmr = (\bmr^X, \bmr^Y)^{\top}$,
where $\bmr^X$ and $\bmr^Y$ are the positions of the subsystems $X$ and $Y$, respectively.
In the measurement process, $Y$ evolves in time depending on $X$ to generate correlations between $X$ and $Y$.
In the feedback process, by contrast, $X$ evolves in time depending on $Y$ to extract work or reduce fluctuations in $X$ by consuming the correlation between $X$ and $Y$.
We write the mean local velocity as $\bmv^{XY}_t(\bmr) = (\bmv^X_t(\bmr), \bmv^Y_t(\bmr))^{\top}$, and the probability distribution of the total system as $p_t^{XY}(\bmr)$.
The thermodynamic costs of the subsystems under information processing are given by the partial entropy productions~\cite{Parrondo:2015aa, PhysRevLett.109.180602, Rosinberg_2016}:
\begin{equation}
\Sigma_\tau^i \coloneqq \frac{1}{D} \int_0^\tau \dd t \int \dd \bmr\, \|\bmv_t^i(\bmr)\|^2 p_t^{XY}(\bmr),\quad (i = X, Y).
    \label{eq:partial_entropy_production}
\end{equation}
By definition, the total entropy production satisfies $\Sigma_\tau^{XY} = \Sigma_\tau^X + \Sigma_\tau^Y$.

We consider the optimization problem of thermodynamic costs of measurement and feedback under the constraint that the mutual information of the final distribution matches a desired value $I_f$.
Although there is some freedom in choosing what to adopt as the thermodynamic cost, such as the partial entropy production or the total entropy production,
we consider a setting in which we first optimize $\Sigma_\tau^X$ (resp.\ $\Sigma_\tau^Y$) and subsequently optimize $\Sigma_\tau^Y$ (resp.\ $\Sigma_\tau^X$) for the measurement (resp.\ feedback) process.
For the measurement process, this optimization can be written as
\begin{align}
    \min_{\substack{\{p_t^{XY}, \bmv_t^{XY}\}_{0\leq t \leq \tau} \in S}} \Sigma_{\tau}^Y[\{p_t^{XY}, \bm{v}_t^{XY}\}_{0 \leq t \leq \tau}],
    \label{eq:optimization_problem_measurement}
\end{align}
where $S$ is the set defined as
\begin{equation}
    S \coloneqq \operatorname*{argmin}_{\substack{\{p_t^{XY}, \bmv_t^{XY}\}_{0\leq t \leq \tau}\, \text{s.t.} \\ p_{0}^{XY} = p_{\ini}^{XY},\, I[p_{\tau}^{XY}] = I_f}} \, \Sigma_\tau^X[\{p_t^{XY}, \bm{v}_t^{XY}\}_{0 \leq t \leq \tau}].
\end{equation}
The corresponding expression for the feedback process can be obtained by exchanging $X$ and $Y$.
In this case, as a result of the optimization, $\bm{v}^X_t = \bm{0}$ (resp.\ $\bm{v}^Y_t = \bm{0}$) holds throughout the process, and therefore $\Sigma_\tau^X = 0$ (resp.\ $\Sigma_\tau^Y = 0$) holds~\cite{PhysRevResearch.7.023159}.
This situation corresponds to the usual case where the marginal probability distribution of the subsystem being measured (resp.\ performing feedback) does not change in time.
Therefore, the optimization problem~\eqref{eq:optimization_problem_measurement} can be rewritten as
\begin{equation}
    \min_{\substack{p_{\fin}\, \text{s.t.}\\ I_{\fin}[p_{\fin}] = I_f}} \min_{\substack{\{p_t^{XY}, \bmv_t^{XY}\}_{0\leq t \leq \tau}\, \text{s.t.} \\ p_{0}^{XY} = p_{\ini}^{XY},\, p_{\tau}^{XY} = p_{\fin}^{XY}}} \left. \Sigma_{\tau}^Y[\{p_t^{XY}, \bm{v}_t^{XY}\}_{0 \leq t \leq \tau}]\right|_{\Sigma_\tau^X = 0}
    \label{eq:optimization_problem_measurement_transformed}
\end{equation}
for the measurement process, where the counterpart for the feedback process follows by exchanging $X$ and $Y$.
In Appendix~\ref{ap_sec:generalization_measurement_feedback}, as a more general situation, we optimize a weighted sum of the partial entropy productions, which includes the total entropy production and the setting of this section as special cases.

Under this setting, we define the thermodynamic costs of measurement and feedback as $\Sigma_{\tau, \text{meas}}^Y \coloneqq \left.\Sigma_\tau^Y\right|_{\Sigma_\tau^X = 0}$ and $\Sigma_{\tau, \text{fb}}^X \coloneqq \left.\Sigma_\tau^X\right|_{\Sigma_\tau^Y = 0}$, respectively.
The corresponding transport maps are given by $\bmmcT_{\text{meas}}(\bmr) = (\bmr^X, \bmmcT_{\text{meas}}^{Y|X}(\bmr^Y; \bmr^X))^\top$ for measurement and $\bmmcT_{\text{fb}}(\bmr) = (\bmmcT_{\text{fb}}^{X|Y}(\bmr^X; \bmr^Y), \bmr^Y)^\top$ for feedback,
where $\bmmcT_{\text{meas}}^{Y|X}$ and $\bmmcT_{\text{fb}}^{X|Y}$  denote the transport maps conditioned on subsystems $X$ and $Y$, respectively.

\begin{figure*}[t]
\centering
\includegraphics[width=1.0\linewidth]{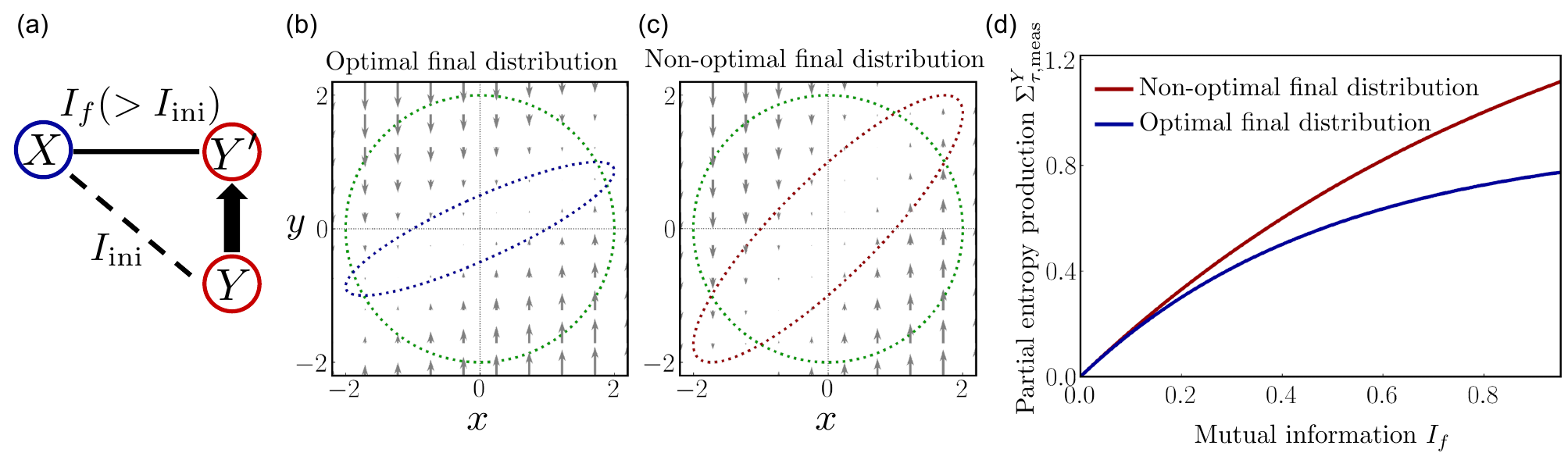}
    \caption{An example of thermodynamically optimal finite-time measurement.
    (a) Schematic illustration of the measurement process.
    The probability current in the $X$ direction vanishes, while the measuring system $Y$ evolves in time to generate a correlation $I_f (> I_\ini)$ between $X$ and $Y$.
    (b), (c) Initial distribution (green dashed ellipse), optimal final distribution (blue dashed ellipse), and non-optimal final distribution (red dashed ellipse).
    We represent each by an ellipse whose semiaxes are twice the standard deviations.
    The gray arrows show the optimal protocol $\bmv_t^{\opt}(\bmr)$ at $t = \tau/2$ scaled by $1/8$.
    As the initial distribution, we take a state with zero mutual information, with covariance components $\Xi^{XX} = \Xi_\ini^{YY} = 1.0$ and $\Xi_\ini^{XY} = \Xi_\ini^{YX} = 0$.
    As the final distribution, we take a state with mutual information $I_f = \ln 2$.
    In this case, the optimal final distribution satisfies Eq.~\eqref{eq:optimal_SigmaYY_meas}.
    As a non-optimal distribution, we set $\left(\Xi_\fin^{YY}\right)^{\mathrm{nonopt}} = \Xi_\ini^{YY} = 1.0$.
    (d) Comparison of the partial entropy production required for the measurement process for the optimal and non-optimal final distributions.
    The larger the desired mutual information $I_f$ of the final distributions, the larger the thermodynamic cost required for measurement, which scales as $\left(\Sigma_{\tau, \mathrm{meas}}^Y\right)^{\opt} \propto 1 - e^{- 2 I_f}$.
    The parameters are set to $D = 1.0$ and $\tau = 1.1$.}
    \label{fig:measurement}
\end{figure*}

In the remainder of this section, we consider only the measurement process.
The equations presented below also hold for the feedback process by exchanging $X$ and $Y$.
In analogy with $\tilde{\Sigma}_{\tau}[\bmmcT]$ introduced in Eq.~\eqref{eq:Sigma_tilde}, we define the corresponding quantity for the partial entropy productions as
\begin{align}
    &\tilde{\Sigma}_{\tau, \text{meas}}^Y[\bmmcT_{\text{meas}}^{Y|X}]\nonumber\\
    &\coloneqq \frac{1}{D \tau} \int \dd \bmr\, \|\bmmcT_{\text{meas}}^{Y|X}(\bmr^Y; \bmr^X) - \bmr^{Y}\|^2 p_{\ini}^{XY}(\bmr).
\end{align}
Meanwhile, a previous study~\cite{PhysRevResearch.7.023159} showed the following relation:
\begin{align}
    & \min_{\substack{\{p_t^{XY}, \bmv_t^{XY}\}_{0\leq t \leq \tau} \\ \text{s.t.}\, p_{0}^{Y|X} = p_{\ini}^{Y|X},\, p_{\tau}^{Y|X} = p_{\fin}^{Y|X}}} \Sigma_{\tau, \text{meas}}^Y[\{p_t^{XY}, \bmv_t^{XY}\}_{0 \leq t \leq \tau}] \nonumber\\
    &=\min_{\bmmcT_{\text{meas}}^{Y|X} \text{ s.t. } \bmmcT_{\text{meas}}^{Y|X} \sharp p_{\ini}^{Y|X} = p_{\fin}^{Y|X}} \tilde{\Sigma}_{\tau, \text{meas}}^Y[\bmmcT_{\text{meas}}^{Y|X}],
\end{align}
where $p^{Y|X}$ is the conditional probability distribution of $Y$ given $X$.
Therefore, by almost the same arguments as in Sec.~\ref{subsec:setup_of_optimization}, the optimization problem for the measurement process in Eq.~\eqref{eq:optimization_problem_measurement} is reduced to
\begin{equation}
    \Sigma_\tau^X = 0
\end{equation}
and
\begin{align}
    \min_{\bmmcT_{\text{meas}}^{Y|X} \text{ s.t. } I_\fin[\bmmcT_{\text{meas}}^{Y|X}] = I_f} \tilde{\Sigma}_{\tau, \text{meas}}^Y[\bmmcT_{\text{meas}}^{Y|X}].
    \label{eq:optimization_problem_measurement_reduced}
\end{align}
The mutual information of the final distribution is given by $I_\fin[p_\fin^{XY}] = \int \dd \bmr\, p_\fin^{XY}(\bmr) \iota_\fin(\bmr)$, where $\iota_\fin(\bmr) \coloneqq \ln [p^{XY}_\fin(\bmr)/(p^X_\fin(\bmr^X) p^Y_\fin(\bmr^Y))]$ is the stochastic mutual information at the final time.
In this case, the Lagrangian is given by $\mathcal{L} = \tilde{\Sigma}_{\tau, \text{meas}}^Y[\bmmcT_{\text{meas}}^{Y|X}] + \lambda (I_\fin[\bmmcT_{\text{meas}}^{Y|X}] - I_f)$.
The variational formula for the partial entropy production with respect to the transport map is then given by
\begin{equation}
    \frac{\delta \tilde{\Sigma}_{\tau, \text{meas}}^Y[\bmmcT_{\text{meas}}^{Y|X}]}{\delta \bmmcT_{\text{meas}}^{Y|X}(\bmr^Y; \bmr^X)} = \frac{2}{D \tau} (\bmmcT_{\text{meas}}^{Y|X}(\bmr^Y; \bmr^X) - \bmr^{Y}) p_{\ini}^{XY}(\bmr).
    \label{eq:variation_partial_EP}
\end{equation}

In general, the minima of the two partial entropy productions are incompatible and exhibit a trade-off.
Previous studies~\cite{PhysRevResearch.7.023159, PhysRevResearch.7.013329} introduced the Pareto front of the partial entropy productions to represent such trade-offs between two incompatible quantities.
The minimization problems considered here correspond to endpoints of the Pareto front in Ref.~\cite{PhysRevResearch.7.023159}, and the optimal transport map at these points is the Knothe--Rosenblatt map~\cite{Villani_optimal_transport, Santambrogio_Optimal_Transport, PhysRevResearch.7.023159}.

\subsubsection{Measurement}\label{subsubsec:measurement}
We consider a process in which the subsystem $Y$ measures $X$ within finite time $\tau$ to generate more mutual information than the initial distribution $I_f>I_\ini$.
Here, we take the thermodynamic cost as $C_\tau = \Sigma_\tau^Y|_{\Sigma_\tau^X = 0}$ ($\tilde{C}_\tau = \tilde{\Sigma}_{\tau, \text{meas}}^Y[\bmmcT_{\text{meas}}^{Y|X}]$),
and the constraint on the final distributions to be $I_\fin[\bmmcT_{\text{meas}}^{Y|X}] = I_f$.
We have to obtain the variational formula for mutual information, which corresponds to the second term in Eq.~\eqref{eq:Lagrangian_variation_transport_map}.
The variation of the mutual information is given by
\begin{equation}
    \frac{\delta I_{\fin}[\bmmcT_{\text{meas}}^{Y|X}]}{\delta \mcT_{\text{meas}, i}^{Y|X}} = \partial_{r_j} \left\{\iota_\fin (\bm{\mathcal{T}_{\text{meas}}}(\bm{r})) \right\}\tilde{J}_{ji}(\bm{r}) p_{\ini}^{XY}(\bm{r}).
    \label{eq:variation_mutual_information}
\end{equation}
See Appendix~\ref{ap_subsec_variation_mutual_information} for details.
Combining this with Eq.~\eqref{eq:variation_partial_EP}, the variational equation to be solved is
\begin{eqnarray}
    &&\frac{2}{D \tau}\left(\mcT_{\text{meas}, i}^{Y|X}(\bmr^Y; \bmr^X) - r_i\right) p_{\ini}^{XY}(\bmr)\nonumber\\
    &&+ \lambda \partial_{r_j} \left\{\iota_\fin (\bm{\mathcal{T}}(\bm{r})) \right\}\tilde{J}_{ji}(\bm{r}) p_{\ini}^{XY}(\bm{r}) = 0.
    \label{eq:variational_eq_measurement}
\end{eqnarray}
Together with the constraint $I_\fin[\bmmcT_{\text{meas}}^{Y|X}] = I_f$, we have the optimal transport map and the optimal final distribution that minimizes the partial entropy production in the measurement process.

\begin{figure*}[tb]
    \centering
    \includegraphics[width=1.0\linewidth]{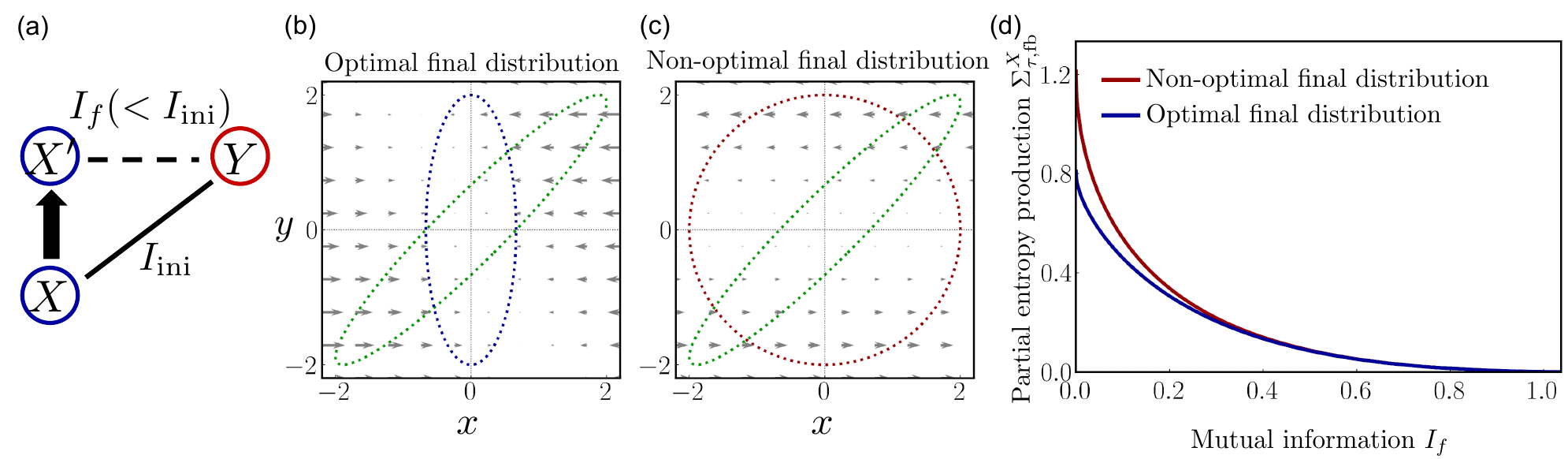}
    \caption{An example of thermodynamically optimal finite-time feedback.
        (a) Schematic illustration of the feedback process.
        The probability current in the $Y$ direction vanishes, while the measured system $X$ evolves in time, consuming the correlation and transitioning to a state with $I_f (< I_\ini)$.
        (b), (c) Initial distribution (green dashed ellipse), optimal final distribution (blue dashed ellipse), and non-optimal final distribution (red dashed ellipse).
        We represent each by an ellipse whose semiaxes are twice the standard deviations.
        The gray arrows show the optimal protocol $\bmv_t^{\opt}(\bmr)$ at $t = \tau/2$ scaled by $1/8$.
        As the initial distribution, we take a state with mutual information $I_\ini = \ln 3$.
        The covariance components are $\Xi_\ini^{XX} = \Xi^{YY} = 1.0$ and $\Xi_\ini^{XY} = \Xi_\ini^{YX} = 2\sqrt{2}/3$.
        As the final distribution, we take a state with mutual information $I_f = 0$.
        In this case, the optimal final distribution satisfies Eq.~\eqref{eq:optimal_SigmaXX_fb}.
        As a non-optimal distribution, we set $\left(\Xi_\fin^{XX}\right)^{\mathrm{nonopt}} = \Xi_\ini^{XX} = 1.0$.
        (d) Comparison of the partial entropy production required for the feedback process for the optimal and non-optimal final distributions.
        The smaller the mutual information $I_f$ of the final distributions, the larger the thermodynamic cost required for feedback.
        The parameters are set to $D = 1.0$ and $\tau = 1.1$.
    }
    \label{fig:feedback}
\end{figure*}

As an illustration, we restrict ourselves to the case of Gaussian distributions throughout the process, which allows us to analytically determine the optimal final distribution and the partial entropy production for the optimal measurement process.
In Fig.~\ref{fig:measurement},
we plot the initial distribution with zero mutual information, the optimal final distribution with mutual information $I_f$ obtained from our framework,
and a fixed non-optimal final distribution with the same $I_f$.
We also compute and plot the partial entropy production $\Sigma_{\tau, \mathrm{meas}}^Y$ for the optimal and non-optimal final distributions.
With the correlation coefficient $\rho \coloneqq \Xi^{XY}/\sqrt{\Xi^{XX} \Xi^{YY}}$, the mutual information is given by $I = - \frac{1}{2} \ln (1 - \rho^2)$,
and thus under the constraint $I_\fin = I_f$ we have $\rho_\fin = \rho_f = \sqrt{1 - e^{- 2 I_f}}$.
Under this condition, the optimal variance of $Y$, $(\Xi^{YY}_{\fin})^{\opt}$, and the optimal value of the partial entropy production required for measurement, $\left(\Sigma_{\tau, \mathrm{meas}}^Y\right)^{\opt}$, are given by
\begin{equation}
    \left(\Xi_\fin^{YY}\right)^{\opt} = (1 - \rho_f^2) \Xi_\ini^{YY},
    \label{eq:optimal_SigmaYY_meas}
\end{equation}
\begin{equation}
    \left(\Sigma_{\tau, \mathrm{meas}}^Y\right)^{\opt} = \frac{\Xi_\ini^{YY} \rho_f^2}{D \tau}.
    \label{eq:optimal_sigmaY_meas}
\end{equation}
See Appendix~\ref{ap_sec:measurement_and_feedback} for details.

\subsubsection{Feedback}
We consider a process in which the subsystem $Y$ performs feedback on $X$ in finite time $\tau$ to consume an amount $I_\ini - I_f > 0$ of mutual information.
Here, we take the thermodynamic cost as $C_\tau = \Sigma_\tau^X|_{\Sigma_\tau^Y = 0}$ ($\tilde{C}_\tau = \tilde{\Sigma}_{\tau, \text{fb}}^X[\bmmcT_{\text{fb}}^{X|Y}]$).
The variational equation to be solved is obtained from Eq.~\eqref{eq:variational_eq_measurement} by exchanging $X$ and $Y$.

As in the measurement case, we restrict ourselves to Gaussian distributions throughout the process.
Figure~\ref{fig:feedback} shows an example of thermodynamically optimal feedback for Gaussian distributions.
We compute and plot the partial entropy production $\Sigma_{\tau, \mathrm{fb}}^X$ for the optimal and non-optimal final distributions.
We plot the initial distribution with mutual information $I_\ini = \ln 3$.
The optimal final distribution and a non-optimal final distribution are also plotted in the case of $I_f=0$.
In this setting, in analogy with the measurement process, we can analytically determine the optimal final distribution and the partial entropy production for the optimal feedback process.
The optimal variance of $X$, $\left(\Xi_\fin^{XX}\right)^{\opt}$, and the optimal value of the partial entropy production required for feedback, $\left(\Sigma_{\tau, \mathrm{fb}}^X\right)^{\opt}$, are given by
\begin{equation}
    \left(\Xi_\fin^{XX}\right)^{\opt} = \left(\rho_\ini \rho_f + \sqrt{(1 - \rho_\ini^2) (1 - \rho_f^2)}\right)^2 \Xi_\ini^{XX},
    \label{eq:optimal_SigmaXX_fb}
\end{equation}
\begin{eqnarray}
    \left(\Sigma_{\tau, \mathrm{fb}}^X\right)^{\opt} &=& \frac{\Xi_\ini^{XX}}{D \tau} \left[1 - \left\{\rho_\ini \rho_f + \sqrt{(1 - \rho_\ini^2) (1 - \rho_f^2)}\right\}^2\right].\nonumber\\
    \label{eq:optimal_sigmaX_fb}
\end{eqnarray}
See Appendix~\ref{ap_sec:measurement_and_feedback} for details.

\section{Conclusion and Discussion}\label{sec:conclusion_and_discussion}
We have constructed a general framework for optimizing the thermodynamic cost required to perform various tasks in overdamped Langevin systems.
That is, we have developed a framework in which quantities depending on the final distributions attain desired values with minimal thermodynamic cost.
To this end, we combined optimal transport theory with a variational method over the transport map.
The standard setup of thermodynamic speed limits based on optimal transport theory fixes the initial and final distributions,
whereas we relax that assumption and instead constrain quantities that depend on the final distributions specific to the thermodynamic task.
Consequently, among the many possible final distributions that realize a given task,
our general framework, Eqs.~\eqref{eq:Lagrangian_variation_transport_map} and~\eqref{eq:constraint}, determines the optimal transport map leading to the optimal final distribution.
In the broad setting of thermodynamic tasks where a functional of the final distribution is constrained to match a desired value, we derived the variational equation for the optimal transport map.

Our framework is applicable to far-from-equilibrium situations, and thus admits various applications (see Table~\ref{table:fixed_quantities_applications}, see also Sec.~\ref{sec:applications}).
In contrast to the case where the final distribution is fixed,
we demonstrated that further optimization of the thermodynamic cost is possible
in analytically tractable examples and in numerical examples.

Thermodynamic speed limits based on optimal transport theory have been extended not only to the overdamped Langevin systems considered in this study~\cite{Aurell:2012aa, PhysRevResearch.3.043093, dechant2019thermodynamicinterpretationwassersteindistance},
but also to Markov jump processes~\cite{PhysRevLett.126.010601, Dechant_2022, PhysRevResearch.5.013017, PhysRevX.13.011013}, open quantum systems~\cite{PhysRevLett.126.010601, PhysRevX.13.011013},
and nonlinear systems~\cite{PhysRevResearch.5.013017, PhysRevResearch.7.033011}.
Extending our framework to these cases is left for future work.
For Markov jump processes, previous studies~\cite{PhysRevResearch.6.033239,td2s-819q} have tackled this problem for measurement and feedback setups, but a unified framework is still elusive.

From an experimental viewpoint, optimizing the partial entropy production generally requires complete control of nonconservative forces in addition to the potential~\cite{PhysRevResearch.7.023159}.
Moreover, thermodynamically optimal information processing in multipartite systems demands flexible control of interactions.
Even in a system with optical tweezers where optimal transport has been experimentally implemented~\cite{Oikawa:2025aa}, arbitrary control of interactions remains challenging.

\begin{acknowledgments}
    This work is supported by JST ERATO Grant No. JPMJER2302, Japan.
    K.T. and R.N. are supported by World-leading Innovative Graduate Study Program for Materials Research, Information, and Technology (MERIT-WINGS) of the University of Tokyo.
    K.F. is supported by JSPS KAKENHI Grant Nos. JP23K13036 and JP24H00831.
    T.S. is also supported by JST CREST Grant No. JPMJCR20C1 and by Institute of AI and Beyond of the University of Tokyo.
\end{acknowledgments}

\appendix
\section{Several remarks}\label{ap_sec:several_remarks}
In this section, we provide several remarks on the general framework presented in Sec.~\ref{sec:general_framework}.

We first comment that the constraints on final distributions can be generalized to inequality constraints $\bmA[\bmmcT] \leq \bmA_f$ or $\bmA[\bmmcT] \geq \bmA_f$.
In this case, the Lagrange multiplier $\bmlambda$ must be nonnegative or nonpositive, respectively, according to the Karush--Kuhn--Tucker conditions~\cite{Clarke:2013aa}.
Conditions of convexity of the Lagrangian density $\mathscr{L}(\bmr, \bmmcT, \pdv{\bmmcT}{\bmr}, \bmlambda)$ may differ from those for equality-constrained problems.

We also comment on the differentiability of the transport map.
In deriving the Euler--Lagrange equation, we assume the differentiability of the transport map.
However, in singular settings such as when the final distribution is concentrated at a single point~\cite{PhysRevLett.125.100602, PhysRevE.102.032105}, the solution of the Euler--Lagrange equation can fail to be differentiable.
In such cases, it is known that singular solutions can be handled by a weak formulation of the Euler--Lagrange equation~\cite{Clarke:2013aa}.

Finally, we discuss the numerical cost of our framework if analytic solutions are unavailable.
Suppose we discretize each axis into $k$ grid points and numerically solve the variational equation by employing Newton's method or one of its variants.
In our case, for $d$-dimensional systems, the number of unknown variables scales as $\mathcal{O}(k^d)$ and the numerical cost of finding a stationary solution is $\mathcal{O}(k^d)$ per iteration.
Therefore, one-dimensional systems can be solved efficiently, as $k \sim 10^4$ is feasible in practice and one can exploit the monotonicity $\mcT'(r) > 0$.
In our numerical example of free energy control, we set $k=601$. For two-dimensional systems, we expect that a discretization with $k \sim 10^2$--$10^3$ is still feasible in practice.

\section{Derivation of the main results}\label{ap_sec:derivation_main_results}
In this section, we provide the details of the derivations of the results in Sec.~\ref{sec:general_framework}.
\subsection{Variation of free energy}\label{ap_subsec:variation_free_energy}
To obtain the variational formula for the work with respect to the transport map in Eq.~\eqref{eq:variation_work}, we need the variation of the nonequilibrium free energy in Eq.~\eqref{eq:free_energy_variation}.
In this subsection, we give the details of that derivation.

For the state $p_{\fin}$ and the potential $V_\tau$ at time $t = \tau$, the nonequilibrium free energy is
\begin{eqnarray}
    \mathcal{F}_{\fin} &=& \beta^{-1} D(p_{\fin} || p_{\fin}^{\mathrm{eq}}) + \mathcal{F}_{\fin}^{\mathrm{eq}}\nonumber\\
                     &=& \beta^{-1} \int \dd \bm{r}\, p_{\fin}(\bm{r}) \ln \frac{p_{\fin}(\bm{r})}{p_{\fin}^{\mathrm{eq}}(\bm{r})} + \mathcal{F}_{\fin}^{\mathrm{eq}}\nonumber\\
                     &=& \beta^{-1} \int \dd \bm{r}\, p_{\ini}(\bm{r}) \ln \frac{p_{\ini}(\bm{r})}{p_{\fin}^{\mathrm{eq}}(\bm{\mathcal{T}}(\bm{r})) \det J_{\bm{\mathcal{T}}}(\bm{r})}\nonumber\\
                     &\phantom{=}& +\mathcal{F}_{\fin}^{\mathrm{eq}},
\end{eqnarray}
where we used Eq.~\eqref{eq:change_variable}.
Thus, we have
\begin{eqnarray}
    \frac{\delta \mcF_\tau}{\delta \mcT_i} = - \beta^{-1} \left(\frac{\delta}{\delta \mcT_i} \int \dd \bmr\, p_{\ini}(\bmr) \ln p_{\fin}^{\mathrm{eq}}(\bmmcT(\bmr))\right.\nonumber\\
    \left. + \frac{\delta}{\delta \mcT_i} \int \dd \bmr\, p_{\ini}(\bmr) \ln \det J_{\bmmcT}(\bmr)\right),
\end{eqnarray}
where the first term is given by
\begin{eqnarray}
    \frac{\delta}{\delta \mcT_i} \int \dd \bmr\, p_{\ini}(\bmr) \ln p_{\fin}^{\mathrm{eq}}(\bmmcT(\bmr))\nonumber\\
     = p_\ini(\bmr) \left. \partial_{r_i'} \ln p_{\fin}^{\mathrm{eq}}(\bmr')\right|_{\bmr' = \bmmcT(\bmr)}.
\end{eqnarray}
On the other hand, the second term requires the variation of $\det J_{\bmmcT}(\bmr)$.
For this we use Jacobi's formula~\cite{Magnus_Matrix_Differential_Calculus}: for a matrix $A(t)$ differentiable with respect to a parameter $t$,
\begin{equation}
    \dv{\det A(t)}{t} = \det A(t) \tr\left(A^{-1}(t) \dv{A(t)}{t}\right).
\label{eq:Jacobi_formula}
\end{equation}
Using Eq.~\eqref{eq:Jacobi_formula}, we have
\begin{eqnarray}
    &&\delta\int \dd \bmr\, p_{\ini}(\bm{r}) \ln \det J_{\bm{\mathcal{T}}}(\bm{r})\nonumber\\
    &&= \int \dd \bmr\, p_{\ini}(\bm{r}) \tr[J_{\bm{\mathcal{T}}}(\bm{r})^{-1} \delta J_{\bmmcT}(\bmr)]\nonumber\\
    &&= \int \dd \bmr\, p_{\ini}(\bmr) \tilde{J}_{ji}(\bmr) \delta \left(\partial_{r_j} \mcT_i(\bmr)\right)\nonumber\\
    &&= - \int \dd \bmr\, \delta \mcT_i \partial_{r_j} \left(p_{\ini}(\bmr) \tilde{J}_{ji}(\bmr)\right),
\end{eqnarray}
that is,
\begin{equation}
    \frac{\delta}{\delta \mcT_i} \int \dd \bmr\, p_{\ini}(\bm{r}) \ln \det J_{\bm{\mathcal{T}}}(\bm{r}) = - \partial_{r_j} \left(p_{\ini}(\bmr) \tilde{J}_{ji}(\bmr)\right).
\end{equation}
Combining the above yields Eq.~\eqref{eq:free_energy_variation}.

\subsection{Variation of a functional of the final distribution}\label{ap_subsec:variation_functional_final_distributions}
We give the details of the derivation of Eq.~\eqref{eq:variation_functional_final_distributions}.
We have
\begin{eqnarray}
    \delta \int \dd \bmr\, \bmg (p_{\fin}(\bmr), \bmr) = \int \dd \bmr\, \left.\pdv{\bmg(X, \bmr)}{X}\right|_{X = p_{\fin}(\bmr)} \delta p_{\fin}(\bmr),\nonumber\\
    \label{eq:variation_functional_final_distributions_delta_p_tau}
\end{eqnarray}
where, using Eq.~\eqref{eq:change_variable},
\begin{equation}
    \delta p_{\fin}(\bmr) = \delta \frac{p_{\ini}(\bm{\mathcal{T}}^{-1}(\bm{r}))}{\det J_{\bm{\mathcal{T}}}(\bm{\mathcal{T}}^{-1}(\bm{r}))}.
\label{eq:variation_p_tau_change_variable}
\end{equation}
We then need to express $\delta \bmmcT^{-1}(\bmr)$ as a variation with respect to $\bmmcT(\bmr)$.
In Sec.~\ref{ap_subsubsec:variation_inverse_function_transport_map} below, we express $\delta \bmmcT^{-1}(\bmr)$ in terms of variations of $\bmmcT(\bmr)$,
from which we obtain the variation of $p_{\fin}(\bmr)$ in Eq.~\eqref{eq:variation_p_tau_change_variable}, and then Eq.~\eqref{eq:variation_functional_final_distributions}.

\subsubsection{Variation of the inverse function of the transport map}\label{ap_subsubsec:variation_inverse_function_transport_map}
Here we derive the variation of $\bmmcT^{-1}(\bmr)$ needed for Eq.~\eqref{eq:variation_p_tau_change_variable}.
For an infinitesimal variation $\delta \bmmcT$ of the transport map,
\begin{equation}
    (\bmmcT + \delta \bmmcT)\left((\bmmcT + \delta \bmmcT)^{-1}(\bmr)\right) = \bmr
    \label{eq:transportmap_variation_identity}
\end{equation}
holds, while
\begin{eqnarray}
    && \bmmcT\left((\bmmcT + \delta \bmmcT)^{-1} (\bmr) \right)\nonumber\\
    &&= \bmmcT\left(\bmmcT^{-1}(\bmr)\right) + J_{\bmmcT}\left(\bmmcT^{-1}(\bmr)\right) \delta \bmmcT^{-1}(\bmr)\nonumber\\
    &&= \bmr + J_{\bmmcT}\left(\bmmcT^{-1}(\bmr)\right) \delta \bmmcT^{-1}(\bmr).
    \label{eq:inv_transportmap_variation_argument}
\end{eqnarray}
Subtracting the two sides of Eqs.~\eqref{eq:transportmap_variation_identity} and~\eqref{eq:inv_transportmap_variation_argument} gives
\begin{equation}
\delta \bmmcT\left(\bmmcT^{-1}(\bmr)\right) = - J_{\bmmcT}\left(\bmmcT^{-1}(\bmr)\right) \delta \bmmcT^{-1}(\bmr).
\end{equation}
Multiplying on the left by $\left(J_{\bmmcT}\left(\bmmcT^{-1}(\bmr)\right)\right)^{-1} = J_{\bmmcT^{-1}}(\bmr)$ yields
\begin{eqnarray}
    \delta \bm{\mcT}^{-1}(\bmr) &=& - \left(J_{\bmmcT}(\bmmcT^{-1}(\bmr))\right)^{-1} \delta \bmmcT\left(\bmmcT^{-1}(\bmr)\right) \nonumber\\
                                &=& - J_{\bmmcT^{-1}}(\bmr) \delta \bmmcT\left(\bmmcT^{-1}(\bmr)\right),
\end{eqnarray}
that is,
\begin{equation}
    \delta \mcT_i^{-1}(\bmr) = - \tilde{J}_{ij}\left(\bmmcT^{-1}(\bmr)\right) \delta \mcT_j\left(\bmmcT^{-1}(\bmr)\right).
    \label{eq:variation_inverse_transport_map_i}
\end{equation}

\subsubsection{Variation of \texorpdfstring{$p_{\fin}(\bmr)$}{p tau (r)}}
Using the result obtained in Sec.~\ref{ap_subsubsec:variation_inverse_function_transport_map}, the variation of $p_{\fin}(\bmr)$ is given by
\begin{eqnarray}
    \delta p_{\fin}(\bmr) &=& \delta \frac{p_{\ini}(\bm{\mathcal{T}}^{-1}(\bm{r}))}{\det J_{\bm{\mathcal{T}}}(\bm{\mathcal{T}}^{-1}(\bm{r}))}\nonumber\\
                        &=& \frac{\delta p_{\ini}(\bm{\mathcal{T}}^{-1}(\bm{r}))}{\det J_{\bm{\mathcal{T}}}(\bm{\mathcal{T}}^{-1}(\bm{r}))}\nonumber\\
                        &\phantom{=}& - \frac{p_{\ini}(\bm{\mathcal{T}}^{-1}(\bm{r})) \delta \det J_{\bm{\mathcal{T}}}(\bm{\mathcal{T}}^{-1}(\bm{r}))}{\left(\det J_{\bm{\mathcal{T}}}(\bm{\mathcal{T}}^{-1}(\bm{r}))\right)^2}\nonumber\\
                        &=& \frac{\left. \partial_{r_i'} p_{\ini}(\bmr') \right|_{\bmr' = \bmmcT^{-1}(\bmr)} \delta \mcT_i^{-1}(\bmr)}{\det J_{\bm{\mathcal{T}}}(\bm{\mathcal{T}}^{-1}(\bm{r}))}\nonumber\\
                        &\phantom{=}& - \frac{p_{\ini}(\bm{\mathcal{T}}^{-1}(\bm{r})) \tr[\left(J_{\bmmcT}\left(\bmmcT^{-1}(\bmr)\right)\right)^{-1}\delta J_{\bm{\mathcal{T}}}(\bm{\mathcal{T}}^{-1}(\bm{r}))]}{\det J_{\bm{\mathcal{T}}}(\bm{\mathcal{T}}^{-1}(\bm{r}))}.\nonumber\\
                        \label{eq:variation_p_tau_Appendix}
\end{eqnarray}
Here, $\delta J_{\bmmcT}(\bmmcT^{-1}(\bmr))$ is given by
\begin{widetext}
\begin{eqnarray}
    \delta \left(J_{\bm{\mathcal{T}}}(\bm{\mathcal{T}}^{-1}(\bm{r}))\right)_{ij} &=& \left. \delta \left(\partial_{r_j'} \mcT_i(\bmr') \right|_{\bmr' = \bmmcT^{-1}(\bmr)}\right)\nonumber\\
                                                                                &=& \left. \partial_{r_j'} (\mcT_i + \delta \mcT_i)(\bmr') \right|_{\bmr' = \left(\bmmcT + \delta \bmmcT \right)^{-1}(\bmr)} - \left. \partial_{r_j'}\mcT_i(\bmr')\right|_{\bmr' = \bmmcT^{-1}(\bmr)}\nonumber\\
                                                                                &=& \left. \partial_{r_j'} \mcT_i(\bmr') \right|_{\bmr' = \left(\bmmcT + \delta \bmmcT \right)^{-1}(\bmr)} - \left. \partial_{r_j'}\mcT_i(\bmr')\right|_{\bmr' = \bmmcT^{-1}(\bmr)} + \left. \partial_{r_j'} \delta \mcT_i(\bmr')\right|_{\bmr' = \bmmcT^{-1}(\bmr)}\nonumber\\
                                                                                &=& \left. \partial_{r_k'} \partial_{r_j'} \mcT_i(\bmr') \right|_{\bmr' = \bmmcT^{-1}(\bmr)} \delta \mcT_k^{-1} (\bmr) + \left. \partial_{r_j'} \delta \mcT_i(\bmr')\right|_{\bmr' = \bmmcT^{-1}(\bmr)}\nonumber\\
                                                                                &=& H_{kji}\left(\bmmcT^{-1}(\bmr)\right) \delta \mcT_k^{-1} (\bmr) + \left. \partial_{r_j'} \delta \mcT_i(\bmr')\right|_{\bmr' = \bmmcT^{-1}(\bmr)},
\end{eqnarray}
where $H_{kji}(\bmr) = \partial_{k} \partial_{r_j} \mcT_i(\bmr)$.
Rewriting $\left(\left(J_{\bmmcT}\left(\bmmcT^{-1}(\bmr)\right)\right)^{-1}\right)_{ji}$ as $\tilde{J}_{ji}\left(\bmmcT^{-1} (\bmr)\right)$ and using Eq.~\eqref{eq:variation_inverse_transport_map_i}, we have
\begin{eqnarray}
        \delta p_{\fin} (\bmr) &=& \frac{1}{\det J_{\bm{\mcT}}(\bm{\mcT}^{-1}(\bmr))} \left[- \left.\partial_{r_i'} p_{\ini}(\bmr')\right|_{\bmr' = \bm{\mcT}^{-1}(\bmr)} \tilde{J}_{ij}\left(\bmmcT^{-1}(\bmr)\right) \delta \mcT_j\left(\bmmcT^{-1}(\bmr)\right) - p_{\ini}(\bm{\mcT}^{-1}(\bmr)) \tilde{J}_{ji}(\bm{\mcT}^{-1}(\bmr))\right.\nonumber\\
                               &\phantom{=}&\left. \cdot \left\{- H_{kji}(\bm{\mcT}^{-1}(\bmr)) \tilde{J}_{kl}(\bm{\mcT}^{-1}(\bmr)) \delta \mcT_l(\bm{\mcT}^{-1}(\bmr)) + \left.\partial_{r_j'} \delta \mcT_i(\bmr')\right|_{\bmr' = \bm{\mcT}^{-1}(\bmr)}\right\}\right].
        \label{eq:variation_p_tau}
\end{eqnarray}
Hence, Eq.~\eqref{eq:variation_functional_final_distributions_delta_p_tau} is transformed into
\begin{eqnarray}
        \delta \int \dd \bmr\, \bmg (p_{\fin}(\bmr), \bmr) &=& \int \dd \bmr\, \frac{\hat{\bmg}(\bmr)}{\det J_{\bm{\mathcal{T}}}(\bm{\mathcal{T}}^{-1}(\bm{r}))} \left[- \left.\partial_{r_i'} p_{\ini}(\bmr')\right|_{\bmr' = \bm{\mcT}^{-1}(\bmr)} \tilde{J}_{ij}\left(\bmmcT^{-1}(\bmr)\right) \delta \mcT_j\left(\bmmcT^{-1}(\bmr)\right) \right. \nonumber\\
                                                           &\phantom{=}&\left. - p_{\ini}(\bm{\mcT}^{-1}(\bmr)) \tilde{J}_{ji}(\bm{\mcT}^{-1}(\bmr)) \left\{- H_{kji}(\bm{\mcT}^{-1}(\bmr)) \tilde{J}_{kl}(\bm{\mcT}^{-1}(\bmr)) \delta \mcT_l(\bm{\mcT}^{-1}(\bmr)) + \left.\partial_{r_j'} \delta \mcT_i(\bmr')\right|_{\bmr' = \bm{\mcT}^{-1}(\bmr)}\right\}\right]\nonumber\\
                                                           &=& \int \dd \bmr\, \hat{\bmg}(\bmmcT(\bmr)) \left[- \partial_{r_i} (p_{\ini}(\bmr)) \tilde{J}_{ij} (\bmr) \delta \mcT_j (\bmr) - p_{\ini}(\bmr) \tilde{J}_{ji}(\bmr) \left\{- H_{kji}(\bmr) \tilde{J}_{kl}(\bmr) \delta \mcT_l(\bmr) + \partial_{r_j} \delta \mcT_i \right\}\right]\nonumber\\
                                                           &=& \int \dd \bmr\, \hat{\bmg}(\bmmcT(\bmr)) \left[\left\{- \partial_{r_j} (p_{\ini}(\bmr)) \tilde{J}_{ji} (\bmr) + p_{\ini}(\bmr) \tilde{J}_{kl}(\bmr) H_{jkl}(\bmr) \tilde{J}_{ji}(\bmr) \right\} \delta \mcT_i - p_{\ini}(\bmr) \tilde{J}_{ji}(\bmr) \partial_{r_j} \delta \mcT_i \right].\nonumber\\
\end{eqnarray}
Using
\begin{eqnarray}
        \tilde{J}_{kl}(\bm{r}) H_{jkl}(\bm{r}) \tilde{J}_{ji}(\bm{r}) &=& \tilde{J}_{kl}(\bm{r}) \left(\partial_{r_j} \partial_{r_k} \mcT_l(\bmr)\right) \tilde{J}_{ji}(\bm{r})\nonumber\\
                                                                    &=& \tilde{J}_{kl}(\bm{r}) \left(\partial_{r_k} J_{lj}(\bmr)\right) \tilde{J}_{ji}(\bm{r})\nonumber\\
                                                                    &=& \partial_{r_k} \left(\tilde{J}_{kl}(\bm{r}) J_{lj}(\bmr) \tilde{J}_{ji}(\bm{r})\right) - \left(\partial_{r_k} \tilde{J}_{kl}(\bm{r})\right) J_{lj}(\bmr) \tilde{J}_{ji}(\bm{r})- \tilde{J}_{kl}(\bm{r}) J_{lj}(\bmr) \left(\partial_{r_k} \tilde{J}_{ji}(\bm{r})\right) \nonumber\\
                                                                    &=& \partial_{r_k} \left(\tilde{J}_{kl}(\bmr) \delta_{li}\right) - \left(\partial_{r_k} \tilde{J}_{kl}(\bm{r})\right) \delta_{li} - \delta_{kj} \left(\partial_{r_k} \tilde{J}_{ji}(\bm{r})\right)\nonumber\\
                                                                    &=& - \partial_{r_j} \tilde{J}_{ji}(\bm{r}),
\end{eqnarray}
for $\bmA[p_\fin] = \int \dd \bmr\, \bmg(p_\fin(\bmr), \bmr)$, we have
\begin{eqnarray}
        \delta \int \dd \bmr\, \bmg (p_{\fin}(\bmr), \bmr) &=& \int \dd \bmr\, \hat{\bmg}(\bmmcT(\bmr)) \left[\left\{- \partial_{r_j} (p_{\ini}(\bmr)) \tilde{J}_{ji} (\bmr) - p_{\ini}(\bmr) \partial_{r_j} \left(\tilde{J}_{ji}(\bmr)\right) \right\} \delta \mcT_i - p_{\ini}(\bmr) \tilde{J}_{ji}(\bmr) \partial_{r_j} \delta \mcT_i \right]\nonumber\\
                                                &=& \int \dd \bmr\, \partial_{r_j} \left\{\hat{\bmg}(\bmmcT(\bmr))\right\} \tilde{J}_{ji}(\bmr) p_{\ini}(\bmr) \delta \mcT_i.
\end{eqnarray}
Therefore, we have Eq.~\eqref{eq:variation_functional_final_distributions}.

Specifically, if $\bmA[p_{\fin}]$ is the expectation value of a physical observable at the final time $\bmA[p_{\fin}] = \expval{\bma}_{\fin} = \int \dd \bmr\, \bma(\bmr) p_{\fin}(\bmr)$, then $\bmg(p_{\fin}(\bmr), \bmr) = \bma(\bmr) p_{\fin}(\bmr)$.
Substituting it into Eq.~\eqref{eq:variation_functional_final_distributions}, we have
\begin{eqnarray}
    &&\partial_{r_j} \left\{\left. \pdv{\left(\bma \left(\bmmcT(\bmr)\right) X\right)}{X}\right|_{X = p_{\fin}\left(\bmmcT(\bmr)\right)} \right\} \tilde{J}_{ji}(\bmr) p_{\ini}(\bmr)\nonumber\\
    &&= \partial_{r_j} \left\{ \bma \left(\bmmcT(\bmr)\right) \right\} \tilde{J}_{ji}(\bmr) p_{\ini}(\bmr)\nonumber\\
    &&= \partial_{r_k'}\left. \bma(\bmr')\right|_{\bmr' = \bmmcT(\bmr)} J_{kj}(\bmr) \tilde{J}_{ji}(\bmr) p_{\ini}(\bmr)\nonumber\\
    &&= \partial_{r_k'}\left. \bma(\bmr')\right|_{\bmr' = \bmmcT(\bmr)} \delta_{ki} p_{\ini}(\bmr)\nonumber\\
    &&= \partial_{r_i'}\left. \bma(\bmr')\right|_{\bmr' = \bmmcT(\bmr)} p_{\ini}(\bmr).
\end{eqnarray}
Thus, we have Eq.~\eqref{eq:variation_expectation_value_physical_quantities}, which was derived directly in the main text.
\begin{figure*}[t]
    \centering
    \includegraphics[width=1.0\linewidth]{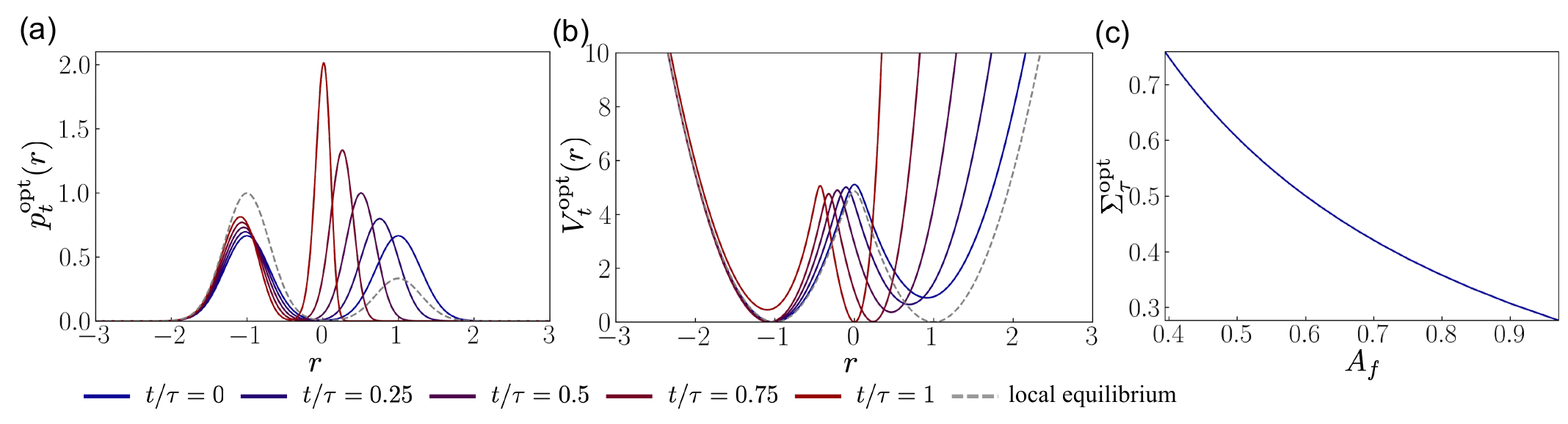}
    \caption{Numerical example of optimal finite-time information erasure.
    (a) Intermediate probability distributions during the optimal erasure process.
    (b) Intermediate optimal potentials during the erasure process.
    The parameter values are $T = 1.0, D = 1.0, \tau = 1.0, m = 1.0, \sigma = 0.3,$ and $\gamma = 2.0$.
    In (a) and (b), we set $A_f \simeq 0.5631$, for which the final erasure error, $\int_0^{\infty} \dd r\, p_{\fin}(r)$, is 0.25.
    In (a), we plot the corresponding local equilibrium distributions $p^{\mathrm{leq}}(r) = 1.5 p_{\ini}(r)\, (r \leq 0)$ and $p^{\mathrm{leq}}(r) = 0.5 p_{\ini}(r)\, (r > 0)$ for comparison (dashed line).
    We fix the constant of the potential in Eq.~\eqref{eq:information_erasure_potential} so that the minimum value of the potential is zero.
    (c) Dependence of the minimal entropy production $\Sigma_\tau^{\opt}$ on the final value of $A[p_{\fin}]$.
    Panel (c) shows that decreasing $A[p_{\fin}]$ increases the minimal entropy production required for erasure.}
    \label{fig:information_erasure}
\end{figure*}
\end{widetext}

\subsection{Variation of mutual information}\label{ap_subsec_variation_mutual_information}
In this subsection, we provide the details of the derivation of Eq.~\eqref{eq:variation_mutual_information}.
We omit the subscript ``meas'' or ``fb'' for simplicity.
The variation of the mutual information at the final time is given by
\begin{eqnarray}
    \delta I_\fin &=& \delta \int \dd \bmr\, p_{\fin}^{XY}(\bmr) \ln \frac{p_{\fin}^{XY}(\bmr)}{p^X_\fin(\bmr^X) p^Y_\fin(\bmr^Y)}\nonumber\\
                  &=& \int \dd \bmr\, \left[\left(\delta p_{\fin}^{XY}(\bmr)\right) \iota_\fin(\bmr) + \delta p_{\fin}^{XY}(\bmr) \right]\nonumber\\
                  &\phantom{=}& - \int \dd \bmr^X\, \delta p_{\fin}^X(\bmr^X) - \int \dd \bmr^Y\, \delta p_{\fin}^Y(\bmr^Y)\nonumber\\
                  &=& \int \dd \bmr\, \left(\delta p_{\fin}^{XY}(\bmr)\right) \iota_\fin(\bmr),
\end{eqnarray}
where
\begin{equation}
\iota_\fin (\bm{r}) = \ln \frac{p^{XY}_{\fin}(\bm{r})}{p^X_\fin(\bm{r}^X) p^Y_\fin(\bm{r}^Y)}
\end{equation}
is the stochastic mutual information at time $t = \tau$.
Using Eq.~\eqref{eq:variation_functional_final_distributions}, we have
\begin{equation}
    \frac{\delta I_\fin}{\delta \mathcal{T}_i} = \partial_{r_j} \left\{\iota_\fin (\bm{\mathcal{T}}(\bm{r})) \right\}\tilde{J}_{ji}(\bm{r}) p_{\ini}^{XY}(\bm{r}).
    \label{eq:variation_mutual_information_Appendix}
\end{equation}
We note that Eq.~\eqref{eq:variation_mutual_information_Appendix} is equivalent to Eq.~\eqref{eq:variation_functional_final_distributions} with $\hat{g}(\bmr) = \iota_\fin(\bmr)$.

\begin{figure*}[ht]
    \centering
    \includegraphics[width = \linewidth]{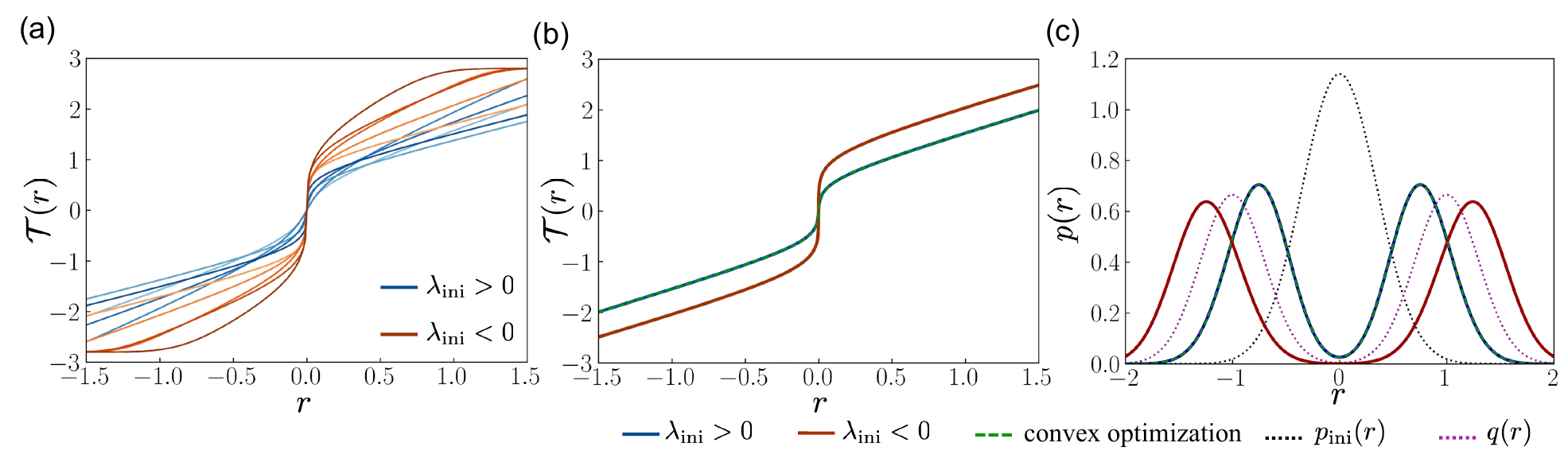}
    \caption{Numerical example of optimal control of free energy in finite time.
    (a) Initial choices of transport map $\mcT(r)$, where blue lines correspond to initial choices of $\lambda_{\ini} > 0$ and red lines correspond to initial choices of $\lambda_{\ini} < 0$.
    (b) Stationary transport maps $\mcT(r)$ for different initial guesses shown in (a) obtained by numerically solving Eq.~\eqref{eq:variational_equation_free_energy_control} together with the constraint on the KL divergence.
    (c) Initial distribution $p_\ini(r)$ (black dotted line), equilibrium distribution with respect to the final potential $q(r)$ (purple dotted line), and optimized final distributions $p_\fin^{\opt}(r)$.
    Initial guesses with $\lambda_{\ini} > 0$ converge to the optimal transport map (green dashed line) which is obtained by solving the convex optimization problem~\eqref{eq:convex_optimization_free_energy_control},
    while initial guesses with $\lambda_{\ini} < 0$ converge to another transport map.
    This indicates that the numerical iteration converges to a stationary solution which is not globally optimal due to the non-convexity for $\lambda_{\ini} < 0$.
    We take the initial distribution as a Gaussian distribution with mean $0$ and standard deviation $0.35$, and the final potential as a negative logarithm of a Gaussian mixture distribution with means $\pm 1$ and standard deviation $0.3$.
    We impose $D_{\mathrm{KL}}(p_\fin || q) = 0.3$, where $q(r)$ is the equilibrium distribution associated with the final potential.
    We set the diffusion coefficient, the inverse temperature and the operation time to $D = 1, \beta = 1,$ and $\tau = 1$.
    As a result, the entropy production for the globally optimal solution $(\lambda_{\ini} > 0)$ is $\Sigma_\tau^{\opt} = 0.249$ and that for the stationary solution $(\lambda_{\ini} < 0)$ is $\Sigma_\tau = 0.923$.
    }
    \label{fig:control_free_energy}
\end{figure*}

\section{An example of information erasure} \label{ap_sec:example_information_erasure}
In this section, we give an example of finite-time optimal information erasure.
We consider a one-dimensional system and take the exponential moment as the quantity to be constrained, which is given by
\begin{equation}
    A[p_{\fin}] = \langle e^{\gamma r} \rangle_{\fin} = \int_{-\infty}^{\infty} \dd r\, e^{\gamma r} p_{\fin}(r).
\end{equation}
We set $\gamma > 0$ and $0 < A_f < A_{\ini} = \int_{-\infty}^{\infty} \dd r\, e^{\gamma r} p_{\ini}(r)$ so that erasure is successful in the sense that the final distribution is localized in the ``0'' state corresponding to the region $r \leq 0$.
This quantity penalizes the final distribution for having a long positive tail at parameter $\gamma$ in the positive $r$ region, which corresponds to the ``1'' state.
We note that we impose a different constraint from the one used in Refs.~\cite{PhysRevLett.125.100602, PhysRevE.102.032105}, where the cumulative distribution function at $r = 0$ was adopted.
We choose the entropy production $\Sigma_\tau$ as the thermodynamic cost $C_\tau$.
We note that, if $\lambda > 0$, the Lagrangian density is convex in $\mcT(r)$, leading to the global optimality of the solution.
The variational formulas corresponding to Eqs.~\eqref{eq:Lagrangian_variation_transport_map} and~\eqref{eq:constraint} are given by
\begin{equation}
    \frac{2}{D \tau} (\mcT(r) - r) p_{\ini}(r) + \lambda \gamma e^{\gamma \mcT(r)} p_{\ini}(r) = 0,
    \label{eq:information_erasure_variation_T}
\end{equation}
\begin{equation}
    \int_{-\infty}^{\infty} \dd r\, e^{\gamma \mcT(r)} p_{\ini}(r) - A_f = 0.
    \label{eq:information_erasure_variation_lambda}
\end{equation}
Equation~\eqref{eq:information_erasure_variation_T} gives the optimal transport map
\begin{equation}
    \mcT^{\opt}(r) = r - \frac{1}{\gamma} W_0(\alpha e^{\gamma r}),
\end{equation}
where $W_0$ is the principal branch of the Lambert $W$ function and we introduce the parameter $\alpha = \lambda D \tau \gamma^2 / 2 > 0$,
which is uniquely determined by solving Eq.~\eqref{eq:information_erasure_variation_lambda}.

In Fig.~\ref{fig:information_erasure}, we demonstrate the finite-time optimal information erasure for this choice of $A[p_{\fin}]$ by numerically determining $\alpha$ from Eq.~\eqref{eq:information_erasure_variation_lambda} and reconstructing the optimal transport map and protocol.
We take the initial distribution to be an equally weighted mixture of two Gaussian distributions
\begin{equation}
    p_{\ini}(r) = \frac{1}{2} \frac{1}{\sqrt{2 \pi \sigma^2}} \left(e^{- \frac{(r - m)^2}{2 \sigma^2}} + e^{- \frac{(r + m)^2}{2 \sigma^2}}\right),
\end{equation}
which is the Boltzmann distribution associated with the double-well potential given by
\begin{align}
    V_0(r) &= - \beta^{-1} \ln p_{\ini}(r) + \mathrm{const.}\nonumber\\
           &= \beta^{-1} \frac{r^2 + m^2}{2 \sigma^2} - \beta^{-1} \ln \cosh\left(\frac{m r}{\sigma^2}\right) + \mathrm{const.}
    \label{eq:information_erasure_potential}
\end{align}
We note that the potential $V_0(r)$ is a double-well potential if $m^2 > \sigma^2$, since the second derivative of $V_0(r)$ at $r = 0$ is given by $V_0''(0) = \beta^{-1} (\sigma^2 - m^2)/\sigma^4$.
In contrast to Refs.~\cite{PhysRevLett.125.100602, PhysRevE.102.032105}, the optimal final distribution is not a singular distribution localized at $r = 0$ but a smooth distribution.

Even for a single thermodynamic task such as information erasure, there is arbitrariness in the choice of the criterion that determines successful achievement of the task.
Our framework can handle general constraints on the final distributions based on theoretical and experimental requirements when performing thermodynamic tasks.

\section{An example of free energy control} \label{ap_sec:example_free_energy_control}

Here, we demonstrate the optimal control of free energy in finite time for a one-dimensional non-Gaussian distribution.
To do so, we fix the potential at $t = \tau$ and impose the constraint that the final distribution $p_\fin$ has a fixed Kullback--Leibler (KL) divergence from the equilibrium distribution with respect to the final potential $q(r) \propto e^{- \beta V_\fin(r)}$, which is equivalent to fixing the free energy change.
Specifically, we fix the equilibrium distribution $q(r)$ as
\begin{equation}
    q(r) = \frac{1}{2} \frac{1}{\sqrt{2 \pi \sigma^2}} \left(e^{- \frac{(r - m)^2}{2 \sigma^2}} + e^{- \frac{(r + m)^2}{2 \sigma^2}}\right),
\end{equation}
which is a mixture of two Gaussian distributions with means $\pm m$ and standard deviation $\sigma$.
Accordingly, the final potential is given by
\begin{align}
    V_{\fin}(r) &= - \beta^{-1} \ln q(r) + \text{const.}\nonumber\\
                &= - \beta^{-1} \ln \left(e^{- \frac{(r - m)^2}{2 \sigma^2}} + e^{- \frac{(r + m)^2}{2 \sigma^2}}\right) + \text{const.}
\end{align}
We consider the optimization problem of minimizing the entropy production $\Sigma_\tau$ under the constraint that the KL divergence is fixed at $D_{\mathrm{KL}}(p_\fin || q) = A_f < D_{\mathrm{KL}}(p_\ini || q)$.
The Lagrangian for this problem is given by
\begin{align}
    \mathcal{L} &= \frac{1}{D \tau} \int \dd r\, (\mcT(r) - r)^2 p_\ini(r) \nonumber\\
                &\phantom{=} + \lambda \left\{\int\dd r\, p_\ini(r) \ln \frac{p_\ini(r)}{q(\mcT(r)) \mcT'(r)} - A_f\right\}.
    \label{eq:Lagrangian_free_energy_control}
\end{align}
We comment on the convexity of the Lagrangian density in Eq.~\eqref{eq:Lagrangian_free_energy_control} with respect to $\mcT(r)$ and $\mcT'(r)$.
The Lagrangian density is not convex in $\mcT(r)$ in general since the convexity is governed by the curvature of $- \lambda \log q$.
For the present Gaussian mixture, $-\log q$ is not convex when $m > \sigma$, since $\left. - \dv[2]{\log q(r)}{r}\right|_{r = 0} = (\sigma^2 - m^2)/\sigma^4 < 0$.
In contrast, the Lagrangian density is convex in $\mcT'(r)$ for $\lambda > 0$ but not for $\lambda < 0$.
As a result, the global optimality of the solution is not guaranteed in general.

Varying the Lagrangian~\eqref{eq:Lagrangian_free_energy_control} with respect to $\mcT(r)$, we obtain the following equation:
\begin{align}
    &\frac{2}{D \tau} (\mcT(r) - r) p_\ini(r) + \lambda \left[- p_\ini(r) \left. \dv{r'} \ln q(r') \right|_{r' = \mcT(r)} \right.\nonumber\\
    &\left. + \dv{r} \left(\frac{p_\ini(r)}{\mcT' (r)}\right)\right] = 0.
    \label{eq:variational_equation_free_energy_control}
\end{align}
We solve this equation numerically and obtain the optimal transport map $\mcT^{\opt}(r)$ and the optimal final distribution $p_\fin$ (see Fig.~\ref{fig:control_free_energy}).
We then check whether the solution of Eq.~\eqref{eq:variational_equation_free_energy_control} matches the solution of the corresponding convex optimization problem in the Kantorovich formulation that guarantees global optimality:
\begin{align}
    &\min_{p_{\fin}\ \mathrm{s. t.}\ D_{\mathrm{KL}}(p_\fin || q) \leq A_f} \min_{\substack{\pi\ \mathrm{s. t.}\\ \int \dd r'\, \pi(r, r') = p_\ini(r),\\ \int \dd r\, \pi(r, r') = p_\fin(r')}}\nonumber\\
    &\int \dd r \dd r'\, \frac{(r' - r)^2}{D \tau} \pi(r, r').
    \label{eq:convex_optimization_free_energy_control}
\end{align}
Here, the global optimality is guaranteed by the convexity of the sublevel set $\{p_\fin\ |\ D_{\mathrm{KL}}(p_\fin || q) \leq A_f\}$, the linearity of the cost function $\int \dd r \dd r'\, \frac{(r' - r)^2}{D \tau} \pi(r, r')$ with respect to $\pi$, and the linearity of the constraint $\int \dd r\, \pi (r, r') = p_\fin(r')$.
The optimal final distribution of Eq.~\eqref{eq:convex_optimization_free_energy_control} satisfies the equality in the constraint $D_{\mathrm{KL}}(p_\fin || q) \leq A_f$, so the solution of the problem in Eq.~\eqref{eq:convex_optimization_free_energy_control} matches the globally optimal solution of the original problem.
The proof of the equality is as follows.
We assume that the optimal final distribution $p_\fin^{\opt}$ satisfies the inequality $D_{\mathrm{KL}}(p_\fin^{\opt} || q) < A_f$.
Then, we can construct a new final distribution $p_{\fin, \varepsilon} = (1 - \varepsilon) p_\fin^{\opt} + \varepsilon p_{\ini}$ for sufficiently small $\varepsilon > 0$ that satisfies the inequality $D_{\mathrm{KL}}(p_{\fin, \varepsilon} || q) < A_f$.
Since $\mathcal{W}_2^2(p_{\ini}, p)$ is convex in $p$, we have $\mathcal{W}_2^2(p_{\ini}, p_{\fin, \varepsilon}) \leq (1 - \varepsilon) \mathcal{W}_2^2(p_{\ini}, p_\fin^{\opt}) + \varepsilon \mathcal{W}_2^2(p_{\ini}, p_{\ini}) = (1 - \varepsilon) \mathcal{W}_2^2(p_{\ini}, p_\fin^{\opt}) < \mathcal{W}_2^2(p_{\ini}, p_\fin^{\opt})$.
This contradicts the optimality of $p_\fin^{\opt}$, and thus the optimal final distribution satisfies the equality in the constraint $D_{\mathrm{KL}}(p_\fin^{\opt} || q) = A_f$.

\section{Details of measurement and feedback}\label{ap_sec:measurement_and_feedback}
In this section, we give the details of the examples treated in Sec.~\ref{subsec:measurement_feedback}.
In both measurement and feedback, we consider two-dimensional systems with coordinates $\bmr = (x, y)^\top$, where $x$ and $y$ represent the degrees of freedom of the $X$ and $Y$ systems, respectively.
Since what is actually transported is the conditional distribution, it suffices to consider one-dimensional Gaussian distributions.
When the initial and final distributions are one-dimensional Gaussians with means $\mu_\ini, \mu_\fin$ and variances $\Xi_\ini, \Xi_\fin$, respectively, the transport map between them is given by~\cite{https://doi.org/10.1002/mana.19901470121}
\begin{equation}
    \mcT(r) = \sqrt{\frac{\Xi_\fin}{\Xi_\ini}}(r - \mu_\ini) + \mu_\fin.
\end{equation}

\subsection{Measurement}
As the initial distribution, we take
\begin{equation}
    p_{\ini}^{XY}(x, y) = \frac{1}{2\pi\sqrt{\Xi^{XX} \Xi_\ini^{YY}}} \exp\left(-\frac{x^2}{2\Xi^{XX}} - \frac{y^2}{2\Xi_\ini^{YY}}\right),
    \label{eq:initial_distribution_measurement}
\end{equation}
which has zero mean and zero mutual information.
Since the measurement does not change the marginal distribution of $X$, the mean and variance of $x$ are fixed at $\mu_t^X = 0$ and $\Xi_t^{XX} = \Xi^{XX}$ for all $t \in [0, \tau]$.
We restrict the final distributions to be Gaussian.
We also adopt an ansatz that the solution to the variational equation~\eqref{eq:variational_eq_measurement} is given by
\begin{equation}
    \mathcal{T}_{\text{meas}}^{Y|X}(y; x) = \sqrt{\frac{\Xi_\fin^{YY}(1 - \rho_\fin^2)}{\Xi_\ini^{YY}}} y + \mu_\fin^Y + \frac{\Xi_\fin^{XY}}{\Xi^{XX}} x.
    \label{eq:ansatz_transport_map_measurement}
\end{equation}
Equation~\eqref{eq:ansatz_transport_map_measurement} is the optimal transport map from the conditional distributions $p_{\ini}^{Y|X}(y|x)$ to $p_{\fin}^{Y|X}(y|x)$,
which are Gaussian distributions with means $\mu_\ini^{Y|X}(x) = 0$ and $\mu_\fin^{Y|X}(x) = \mu_\fin^Y + \frac{\Xi_\fin^{XY}}{\Xi^{XX}} x$, and variances $\Xi_\ini^{Y|X} = \Xi_\ini^{YY}$ and $\Xi_\fin^{Y|X} = \Xi_\fin^{YY}(1 - \rho_\fin^2)$, respectively~\cite{https://doi.org/10.1002/mana.19901470121}.

In this case, $\tilde{J}_{11}(\bmr) = 1$, $\tilde{J}_{12}(\bmr) = 0$, $\tilde{J}_{21}(\bmr) = - \frac{\Xi_\fin^{XY}}{\Xi^{XX}} \sqrt{\frac{\Xi_\ini^{YY}}{\Xi_\fin^{YY}(1 - \rho_\fin^2)}}$, and $\tilde{J}_{22} = \sqrt{\frac{\Xi_\ini^{YY}}{\Xi_\fin^{YY}(1 - \rho_\fin^2)}}$.
Since
\begin{widetext}
\begin{eqnarray}
    \pdv{y} \left(\iota_\fin(\bmmcT_{\text{meas}}(\bmr))\right) &=& \pdv{y} \ln \frac{p_\fin^{XY}(x, \mcT_{\text{meas}}^{Y|X}(y; x))}{p_\fin^{X}(x) p_\fin^Y(\mcT_{\text{meas}}^{Y|X}(y; x))}\nonumber\\
                                                                &=& \pdv{y} \ln \frac{p_\ini(x, y)}{p^X(x) \pdv{\mcT_{\text{meas}}^{Y|X}(y; x)}{y} p_\fin^Y(\mcT_{\text{meas}}^{Y|X}(y; x))} \nonumber\\
                                                                &=& - \frac{1}{\Xi_\ini^{YY}}y + \frac{1}{\Xi_\fin^{YY}}\sqrt{\frac{\Xi_\fin^{YY}(1 - \rho_\fin^2)}{\Xi_\ini^{YY}}} \left(\mcT_{\text{meas}}^{Y|X}(y; x) - \mu_\fin^Y\right),
\end{eqnarray}
we have
\begin{eqnarray}
    &&\frac{2}{D \tau}\left(\sqrt{\frac{\Xi_\fin^{YY}(1 - \rho_\fin^2)}{\Xi_\ini^{YY}}}y + \mu_\fin^Y + \frac{\Xi_\fin^{XY}}{\Xi^{XX}}x - y\right)\nonumber\\
    && + \lambda \sqrt{\frac{\Xi_\ini^{YY}}{\Xi_\fin^{YY} (1 - \rho_\fin^2)}} \left[-\frac{1}{\Xi_\ini^{YY}}y + \frac{1}{\Xi_\fin^{YY}} \sqrt{\frac{\Xi_\fin^{YY}(1 - \rho_\fin^2)}{\Xi_\ini^{YY}}}\left(\sqrt{\frac{\Xi_\fin^{YY}(1 - \rho_\fin^2)}{\Xi_\ini^{YY}}}y + \frac{\Xi_\fin^{XY}}{\Xi^{XX}}x\right) \right] = 0.
\end{eqnarray}
By setting the coefficients of $x$, $y$, and the constant term to zero, we have
\begin{equation}
    \frac{2}{D\tau} \frac{\Xi_\fin^{XY}}{\Xi^{XX}} + \lambda \frac{\Xi_\fin^{XY}}{\Xi^{XX}\Xi_\fin^{YY}} = 0,
\end{equation}
\begin{equation}
    \frac{2}{D\tau} \left(\sqrt{\frac{\Xi_\fin^{YY}(1 - \rho_\fin^2)}{\Xi_\ini^{YY}}} - 1\right) + \lambda \left(- \frac{1}{\Xi_\ini^{YY}}\sqrt{\frac{\Xi_\ini^{YY}}{\Xi_\fin^{YY} (1 - \rho_\fin^2)}} + \frac{1}{\Xi_\fin^{YY}} \sqrt{\frac{\Xi_\fin^{YY}(1 - \rho_\fin^2)}{\Xi_\ini^{YY}}}\right) = 0,
\end{equation}
\begin{equation}
    \frac{2}{D\tau} \mu_\fin^Y = 0.
\end{equation}
Thus, we have
\begin{equation}
    \mu_\fin^Y = 0,
\end{equation}
\begin{equation}
    \Xi_\fin^{YY} = (1 - \rho_\fin^2)\Xi_\ini^{YY},
\end{equation}
\begin{equation}
    \lambda = - \frac{2}{D\tau} \Xi_\fin^{YY}.
\end{equation}
Substituting $\rho_\fin = \rho_f$ obtained from the constraint yields the optimal variance of $Y$ in the final distributions
\eqref{eq:optimal_SigmaYY_meas}, and the partial entropy production required for measurement~\eqref{eq:optimal_sigmaY_meas}.
\subsection{Feedback}
As the initial distribution, we take
\begin{equation}
    p_{\ini}^{XY}(x, y) = \frac{1}{2\pi \sqrt{\left(\Xi_\ini^{XX} \Xi^{YY} - \left(\Xi_\ini^{XY}\right)^2\right)}} \exp\left(- \frac{1}{2} \begin{pmatrix}
    x & y
    \end{pmatrix} \begin{pmatrix}
    \Xi_\ini^{XX} & \Xi_\ini^{XY}\\
    \Xi_\ini^{XY} & \Xi^{YY}
    \end{pmatrix}^{-1} \begin{pmatrix}
    x\\
    y
    \end{pmatrix}\right),
    \label{eq:initial_distribution_feedback}
\end{equation}
which has zero mean and finite mutual information if $\Xi_{\ini}^{XY} > 0$.
Since ideal feedback does not change the marginal distribution of $Y$, the mean and variance of $y$ are fixed at $\mu_t^Y = 0$ and $\Xi_t^{YY} = \Xi^{YY}$ for all $t \in [0, \tau]$.
As in the case of measurement, we restrict the final distributions to be Gaussian.
We also assume that the solution to the variational equation is given by the ansatz:
\begin{equation}
    \mathcal{T}^{X|Y}(x; y) = \sqrt{\frac{\Xi_\fin^{XX}(1 - \rho_\fin^2)}{\Xi_\ini^{XX} (1 - \rho_\ini^2)}}\left(x - \frac{\Xi^{XY}_\ini}{\Xi^{YY}}y\right) + \mu_\fin^X + \frac{\Xi_\fin^{XY}}{\Xi^{YY}} y,
\end{equation}
which is the optimal transport map from the conditional distributions $p_{\ini}^{X|Y}(x|y)$ to $p_{\fin}^{X|Y}(x|y)$, which are Gaussian distributions with means $\mu_\ini^{X|Y}(y) = \frac{\Xi_\ini^{XY}}{\Xi^{YY}} y$ and $\mu_\fin^{X|Y}(y) =  \mu_\fin^X + \frac{\Xi_\fin^{XY}}{\Xi^{YY}} y$, and variances $\Xi_\ini^{X|Y} = \Xi_\ini^{XX} (1 - \rho_\ini^2)$ and $\Xi_\fin^{X|Y} = \Xi_\fin^{XX} (1 - \rho_\fin^2)$, respectively.
Using these, and solving the equation obtained by exchanging $X$ and $Y$ in Eq.~\eqref{eq:variational_eq_measurement} in the same manner as in the measurement setting, we have
\begin{equation}
    \mu_\fin^X = 0,
\end{equation}
\begin{equation}
    \Xi_\fin^{XX} = \left(\rho_\ini \rho_\fin + \sqrt{(1 - \rho_\ini^2) (1 - \rho_\fin^2)}\right)^2 \Xi_\ini^{XX}.
\end{equation}
Substituting $\rho_\fin = \rho_f$ from the constraint yields the optimal variance of $X$ in the final distributions,~\eqref{eq:optimal_SigmaXX_fb},
and the partial entropy production required for feedback,~\eqref{eq:optimal_sigmaX_fb}.
\end{widetext}

\section{Measurement and feedback minimizing a weighted sum of partial entropy productions}\label{ap_sec:generalization_measurement_feedback}
\begin{figure*}[ht]
    \centering
    \includegraphics[width=1.0\linewidth]{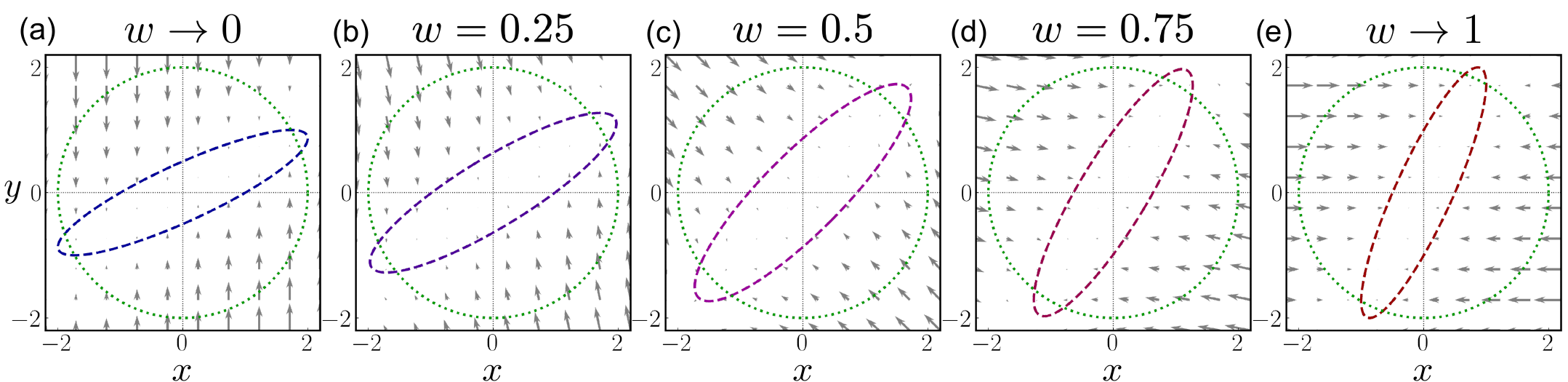}
    \caption{An example of thermodynamically optimal finite-time measurement that minimizes a weighted sum of partial entropy productions.
    In (a)--(e), we plot the initial distribution (green dashed ellipse) and the optimal final distribution (dashed ellipse) for different values of the weight $w$.
    The cases of $w \to 0$ and $w \to 1$ correspond, respectively, to the idealized measurement and feedback discussed in the main text, since in these limits the optimal transport map for the weighted cost converges to the Knothe--Rosenblatt map~\cite{Santambrogio_Optimal_Transport, PhysRevResearch.7.023159}.
    Each distribution is represented by an ellipse whose semiaxes are twice the standard deviations.
    The gray arrows indicate the optimal protocol $\bmv_t^{\opt}(\bmr)$ at $t = \tau/2$ scaled by $1/8$.
    As the initial distribution, we take a state with zero mutual information, with covariance components $\Xi^{XX}_{\ini} = \Xi_\ini^{YY} = 1.0$ and $\Xi_\ini^{XY} = \Xi_\ini^{YX} = 0$.
    As the final distribution, we take a state with mutual information $I_f = \ln 2$.
    The parameters are set to $D = 1.0$ and $\tau = 1.1$.}
    \label{fig:weighted_measurement}
\end{figure*}

In Sec.~\ref{subsec:measurement_feedback}, we considered measurement and feedback processes in which we first optimize one partial entropy production and subsequently optimize the other.
In this section, we optimize a weighted sum of the partial entropy productions, which includes the total entropy production and the setting of Sec.~\ref{subsec:measurement_feedback} as special cases.

For a weight $w \in (0, 1)$, we define the weighted sum of the partial entropy productions, $\Sigma_{\tau}^{w} \coloneqq (1 - w) \Sigma_{\tau}^X + w \Sigma_{\tau}^Y (0 < w < 1)$ and use it as the thermodynamic cost.
In particular, $\Sigma_{\tau}^{1/2}$ is equal to one half of the total entropy production.
For a fixed final distribution, the limiting problems of minimizing $\Sigma_{\tau}^{w}$ as $w \to 0$ and $w \to 1$ are equivalent to the idealized measurement and feedback processes in the main text, respectively~\cite{PhysRevResearch.7.023159}.

We consider the following optimization problem:
\begin{equation}
    \min_{p_{\fin}\, \text{s.t.}\, I[p_{\fin}] = I_f} \min_{\substack{\{p_t, \bmv_t\}_{0\leq t \leq \tau} \\ \text{s.t.}\, p_{0} = p_{\ini},\, p_{\tau} = p_{\fin}}} \Sigma_\tau^w[\{p_t, \bmv_t\}_{0 \leq t \leq \tau}].
    \label{eq:optimization_problem_weighted}
\end{equation}
That is, we seek the minimum value of $\Sigma_{\tau}^{w}$ required to prepare a final distribution with mutual information $I_f$, together with the optimal protocol that attains this minimum.
Since the mean local velocities of both $X$ and $Y$ may be nonzero, the distinction between the system of interest and the memory can become ambiguous.
The weighted sum of the partial entropy productions can be written as
\begin{equation}
    \Sigma_{\tau}^{w} = \int_0^\tau \dd t \int \dd \bm{r}\, \bm{v}_t^{XY}(\bm{r})^\top D_{w}^{-1} \bm{v}_t^{XY}(\bm{r}) p_t^{XY}(\bm{r}),
    \label{eq:partial_EP_weighted_sum}
\end{equation}
where
\begin{equation}
    D_{w} \coloneqq \left(\frac{D}{1 - w} I_d\right) \oplus \left(\frac{D}{w} I_d\right).
\end{equation}
Thus, Eq.~\eqref{eq:partial_EP_weighted_sum} can be interpreted as the total entropy production of a system with diffusion coefficients $D/(1 - w)$ and $D/w$ for $X$ and $Y$, respectively.

To put the weighted sum in Eq.~\eqref{eq:partial_EP_weighted_sum} into a more tractable form, we introduce the rescaled coordinates
\begin{equation}
    S_w \coloneqq (\sqrt{1 - w} I_d) \oplus (\sqrt{w} I_d),
\end{equation}
\begin{equation}
    \tilde{\bm{r}} \coloneqq S_w \bm{r}.
\end{equation}
Then, the probability distribution in the $\tilde{\bm{r}}$ coordinates is given by
\begin{align}
    \begin{split}
        \tilde{p}_t^{XY}(\tilde{\bm{r}}) &= \frac{1}{\det S_w} p_t^{XY}(S_w^{-1} \tilde{\bm{r}})\\
                                         &= \frac{1}{[w(1-w)]^{d/2}} p_t^{XY}(\bm{r}).
    \end{split}
\end{align}
Under this transformation, the Fokker--Planck equation becomes
\begin{equation}
    \pdv{\tilde{p}_t^{XY}(\tilde{\bm{r}})}{t} = - \tilde{\nabla}^\top (S_{w} \bm{v}_t^{XY}(S_w^{-1} \tilde{\bm{r}}) \tilde{p}_t^{XY}(\tilde{\bm{r}})).
\end{equation}
Here, $\tilde{\nabla} \coloneqq S_w^{-1} \nabla$ denotes the gradient with respect to $\tilde{\bm{r}}$.
Defining $\tilde{\bm{v}}_t^{XY}(\tilde{\bm{r}}) \coloneqq S_w \bm{v}_t^{XY}(S_w^{-1} \tilde{\bm{r}})$, we obtain
\begin{equation}
    \pdv{\tilde{p}_t^{XY}(\tilde{\bm{r}})}{t} = - \tilde{\nabla}^\top (\tilde{\bm{v}}_t^{XY}(\tilde{\bm{r}}) \tilde{p}_t^{XY}(\tilde{\bm{r}})).
\end{equation}
Furthermore, Eq.~\eqref{eq:partial_EP_weighted_sum} becomes
\begin{equation}
    \Sigma_{\tau}^w = \frac{1}{D} \int_0^\tau \dd t \int \dd \tilde{\bm{r}}\, \|\tilde{\bm{v}}_t^{XY}(\tilde{\bm{r}}) \|^2 \tilde{p}_t^{XY}(\tilde{\bm{r}}),
\end{equation}
which reduces the problem to the form of the total entropy production with diffusion coefficient $D$.

We next express the mutual information in the rescaled coordinates.
The marginal probability distribution of $X$ is given by
\begin{align}
    \begin{split}
        \tilde{p}_t^X(\tilde{\bm{r}}^X) &= \int \dd \tilde{\bm{r}}^Y\, \tilde{p}_t^{XY}(\tilde{\bm{r}}^X, \tilde{\bm{r}}^Y)\\
                                        &= \frac{1}{(1-w)^{d/2}} p_t^X (\sqrt{1-w}^{-1} \tilde{\bm{r}}^X),
    \end{split}
\end{align}
and similarly,
\begin{equation}
    \tilde{p}_t^Y(\tilde{\bm{r}}^Y) = \frac{1}{w^{d/2}} p_t^Y (\sqrt{w}^{-1} \tilde{\bm{r}}^Y).
\end{equation}
Thus, we obtain
\begin{equation}
    \frac{\tilde{p}_t^{XY}(\tilde{\bm{r}}^X, \tilde{\bm{r}}^Y)}{\tilde{p}_t^X(\tilde{\bm{r}}^X) \tilde{p}_t^Y(\tilde{\bm{r}}^Y)} = \frac{p_t^{XY}(\bm{r}^X, \bm{r}^Y)}{p_t^X(\bm{r}^X) p_t^Y(\bm{r}^Y)}.
\end{equation}
Consequently, the mutual information is invariant under the coordinate transformation:
\begin{align}
    \begin{split}
        I &= \int \dd \bm{r}^X \dd \bm{r}^Y\, p_t^{XY}(\bm{r}^X, \bm{r}^Y) \ln \frac{p_t^{XY}(\bm{r}^X, \bm{r}^Y)}{p_t^X(\bm{r}^X) p_t^Y(\bm{r}^Y)}\\
          &= \int \dd \tilde{\bm{r}}^X \dd \tilde{\bm{r}}^Y\, \tilde{p}_t^{XY}(\tilde{\bm{r}}^X, \tilde{\bm{r}}^Y) \ln \frac{\tilde{p}_t^{XY}(\tilde{\bm{r}}^X, \tilde{\bm{r}}^Y)}{\tilde{p}_t^X(\tilde{\bm{r}}^X) \tilde{p}_t^Y(\tilde{\bm{r}}^Y)}.
    \end{split}
\end{align}

In analogy with Eq.~\eqref{eq:Sigma_tilde}, we define
\begin{equation}
    \tilde{\Sigma}_{\tau}^{w}[\tilde{\bmmcT}] \coloneqq \frac{1}{D\tau} \int \dd \tilde{\bm{r}}\, \|\tilde{\mathcal{T}}(\tilde{\bm{r}}) - \tilde{\bm{r}} \|^2 \tilde{p}_{\ini}^{XY}(\tilde{\bm{r}}).
\end{equation}
Then, as in Eq.~\eqref{eq:identity_Sigma_tilde_Sigma}, we have
\begin{align}
    &\min_{\substack{\{p_t, \bm{v}_t\}_{0\leq t \leq \tau} \\ \text{s.t.}\, p_{0} = p_{\ini},\, p_{\tau} = p_{\fin}}} \Sigma_\tau^w [\{p_t, \bm{v}_t\}_{0 \leq t \leq \tau}]\nonumber\\
    &= \min_{\tilde{\bmmcT}\, \text{s.t.}\, \tilde{\bmmcT}_{\sharp} \tilde{p}_{\ini} = \tilde{p}_{\fin}} \tilde{\Sigma}_\tau^{w}[\tilde{\bmmcT}].
\end{align}
The variation of $\tilde{\Sigma}^{w}_{\tau}$ with respect to $\tilde{\bmmcT}$ is given by
\begin{equation}
    \frac{\delta \tilde{\Sigma}_{\tau}^{w}[\tilde{\bmmcT}]}{\delta \tilde{\bmmcT}(\tilde{\bm{r}})} = \frac{2}{D\tau} (\tilde{\mathcal{T}}(\tilde{\bm{r}}) - \tilde{\bm{r}}) \tilde{p}_{\ini}^{XY}(\tilde{\bm{r}}).
    \label{eq:variation_tilde_Sigma}
\end{equation}
The variation of the final mutual information is
\begin{equation}
    \frac{\delta I_{\mathrm{fin}}[\tilde{\bmmcT}]}{\delta \tilde{\mcT}_i (\tilde{\bm{r}})} = \partial_{\tilde{r}_j} \{\tilde{\iota}_{\mathrm{fin}} (\tilde{\bmmcT}(\tilde{\bm{r}}))\} \tilde{J}_{ji} (\tilde{\bmr}) \tilde{p}_{\ini}^{XY}(\tilde{\bm{r}}),
    \label{eq:variation_tilde_I}
\end{equation}
where
\begin{equation}
    \tilde{\iota}_{\mathrm{fin}}(\tilde{\bm{r}}) = \ln \frac{\tilde{p}_{\fin}^{XY}(\tilde{\bm{r}})}{\tilde{p}_{\fin}^X(\tilde{\bm{r}}^X) \tilde{p}_{\fin}^Y(\tilde{\bm{r}}^Y)},
\end{equation}
and
\begin{equation}
    \tilde{J}_{ji}(\tilde{\bm{r}}) = \left. \partial_{\tilde{r}_i'} \tilde{\mathcal{T}}_j^{-1}(\tilde{\bm{r}}') \right|_{\tilde{\bm{r}}' = \tilde{\bmmcT}(\tilde{\bm{r}})}.
\end{equation}
Together with the constraint $I_{\mathrm{fin}}[\tilde{\bmmcT}] = I_f$,
Eqs.~\eqref{eq:variation_tilde_Sigma} and~\eqref{eq:variation_tilde_I} determine the optimal transport map $\tilde{\bmmcT}_{\mathrm{opt}}$ and the corresponding value $\Sigma_{\tau}^{w, \opt}$.
The protocol in the original coordinate system $\bmr$ is obtained by transforming the optimal quantities in the $\tilde{\bmr}$ coordinates back to the $\bmr$ coordinates.

As an example, we consider an optimal measurement process analogous to that in Sec.~\ref{subsubsec:measurement} but minimizing $\Sigma_{\tau}^{w}$ with $w = 0, 0.25, 0.5, 0.75$, and $1$.
Restricting the distributions to be Gaussian throughout the process, we obtain the optimal final distributions as shown in Fig.~\ref{fig:weighted_measurement}.
We see that the optimal final distributions, together with the optimal protocols, depend on the weight $w$.

The infimum of the optimization problem for measurement in Eqs.~\eqref{eq:optimization_problem_measurement} and~\eqref{eq:optimization_problem_weighted} can become zero if arbitrary final distributions are allowed, because strong correlations can then be created on infinitesimally small length scales.
Therefore, the class of admissible probability distributions needs to be restricted.
A physically natural way to do so is to require that the probability distributions vary only on length scales where the overdamped description is valid.
Another way, which we adopt in the examples, is to restrict the admissible probability distributions to Gaussian distributions, for which the infimum is nonzero.
\bibliography{ref}
\end{document}